\documentstyle[aps,psfig]{revtex}

\def\DO{$D^{0}$\ }
\def\dkpi{${D^{0}\ {\rightarrow}\ K\pi} $}
\def\dmumu{${D^{0}\ {\rightarrow}\ \mu^{+}\mu^{-}} $}
\def\dmue{${D^{0}\ {\rightarrow}\ \mu^{\pm}e^{\mp}} $}
\def\dee{${D^{0}\ {\rightarrow}\ e^{+}e^{-}} $}
\def\dll{{D^{0}\ {\rightarrow}\ l^{+}l^{-}}} 
\def\chix{$\chi^{2}/{\rm degree\ of\ freedom}$}
\draft		
\begin{document}
\title{
\begin{flushright}{\normalsize LBNL-43414 \\
\vspace{-0pt}                 FERMILAB-Pub-99/152-E \\
\vspace{-0pt}                 LA-UR-99-2892} \\
\vspace{24pt}
\end{flushright}
Search for flavor-changing neutral currents and
lepton-family-number violation in two-body $D^0$ decays
}

\author{D. Pripstein\cite{DPaddress},
        G. Gidal, P.M. Ho\cite{PMHaddress}, 
        M.S. Kowitt\cite{MSKaddress}, K.B. Luk}
\address{Physics Division, Lawrence Berkeley Laboratory and Department of 
Physics, University of California, Berkeley, California 94720}

\author{L.D. Isenhower, M.E. Sadler, R. Schnathorst\cite{RSaddress}}
\address{Abilene Christian University, Abilene, Texas 79699}

\author{L.M. Lederman\cite{LLaddress}, M.H. Schub\cite{MHSaddress}}
\address{University of Chicago, Chicago, Illinois 60637}

\author{C.N. Brown, W.E. Cooper, K.N. Gounder\cite{KGaddress},
 C.S. Mishra}
\address{Fermi National Accelerator Laboratory, Batavia, Illinois 60510}

\author{T.A. Carey, D.M. Jansen\cite{DMJaddress},
        R.G. Jeppesen\cite{RJaddress},
        J.S. Kapustinsky, \\
        D.W. Lane\cite{DLaddress},
        M.J. Leitch, J.W. Lillberg, P.L. McGaughey, J.M. Moss, J.C. Peng}
\address{Los Alamos National Laboratory, Los Alamos, New Mexico 87545 }

\author{D.M. Kaplan\cite{LLaddress}, W.R. Luebke\cite{LLaddress}, 
        R.S. Preston, 
        J. Sa, V. Tanikella }
\address{Northern Illinois University, DeKalb, Illinois 60115}

\author{R.L. Childers, C.W. Darden, J.R. Wilson}
\address{University of South Carolina, Columbia, South Carolina 29208}

\author{G.C. Kiang, P.K. Teng}
\address{Institute of Physics, Academia Sinica, Taipei, Taiwan}

\author{Y.C. Chen\cite{YCChen}}
\address{National Cheng Kung University, Tainan, Taiwan}

\date{\today}

\maketitle

\begin{abstract}

We present the results of a search for the three neutral charm decays,
\dmue, \dmumu, and \dee.  This study was based on data collected in
Experiment 789 at the Fermi National Accelerator Laboratory using 
800 GeV/$c$ proton-Au and proton-Be interactions.
No evidence is found for any of the decays. Upper limits on
the branching ratios, at the 90\% confidence level, of 
$1.56\times 10^{-5}$ for \dmumu, $8.19\times 10^{-6}$
for \dee\, and $1.72\times 10^{-5}$ for \dmue\ are obtained.

\end{abstract}


\section{INTRODUCTION}

In the Standard Model, the flavor-changing neutral-current decays 
\dee\ and \dmumu\ are forbidden at tree level.\footnote{In this paper, 
the symbol $D^{0}$ denotes both $D^{0}$ and $\overline{D}^{0}$ mesons.} 
At the one-loop level, the decays are GIM- and helicity-suppressed.  
Thus the branching ratios are expected to be very small.  
Lepton-family-number violating decays such as \dmue\ are 
strictly forbidden.
However, extensions of the Standard Model allow
for both flavor-changing neutral currents and lepton-family-number
violation, so detection of such dilepton decays could be taken as
evidence for new physics~\cite{Wyler:1986,Babu:1988,Pakvasa:1994}.
Previous experiments~\cite{Cleo:1996,Beatrice:1997,E771:1996} have
quoted limits of order $10^{-4}$ to $10^{-6}$ for several such decays. 
In this paper 
we describe an experiment that places limits between a few times 
$10^{-5}$ and $10^{-6}$
on the branching ratios for the decays \dmumu,~\dee, and \dmue.

\section{The E789 Spectrometer}

Experiment 789 was carried out in the Meson East beam line at the Fermi
National Accelerator Laboratory, where a beam of 800 GeV/$c$ protons was 
delivered to a fixed target of either gold or beryllium.  

\begin{figure}[hbt]
\centerline{\psfig{figure=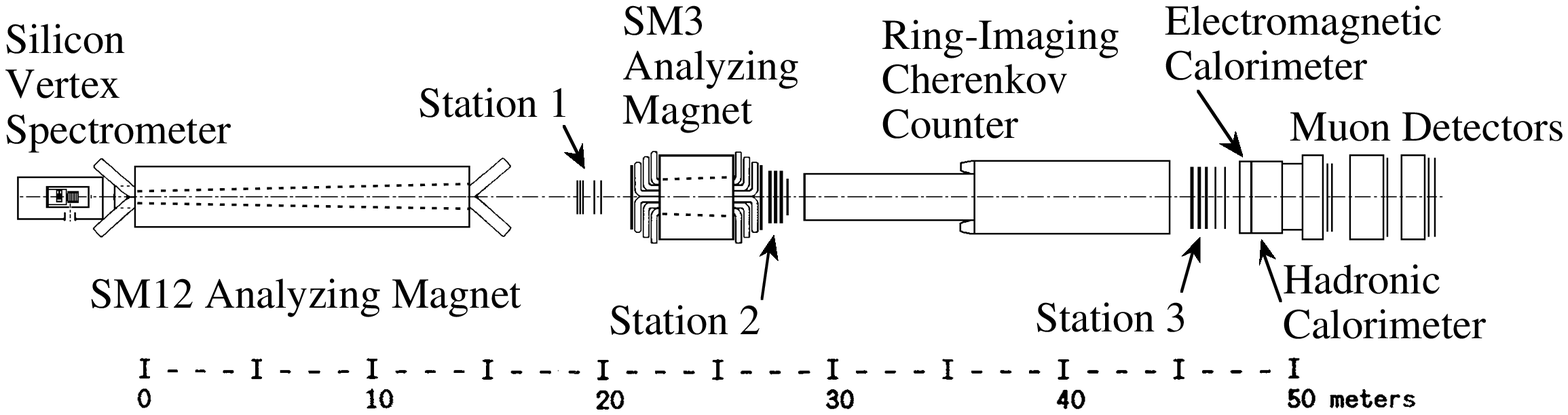,width=6in}}
\caption{Plan view of E789 spectrometer.\label{fig:spectrometer}}
\end{figure}

The spectrometer, shown in Figure~\ref{fig:spectrometer}, 
was optimized for two-body final states with pair rapidity near zero in
the center-of-mass system.
Its main components were
a silicon-strip vertex detector (SSD)
just after the target, a copper beam dump, 
two dipole bending magnets (SM12 and SM3),
three stations of drift chambers and 
hodoscopes, a sampling calorimeter, and a muon-identification station located 
at the end of the spectrometer. 
The ring-imaging Cherenkov detector was 
not used in this analysis.

\subsection{Target}
The target apparatus was installed in the beam vacuum. 
Table~\ref{target_table} gives the dimensions of the targets used
for this analysis.  
The targets were much wider than the beam in the $x$ (horizontal)
dimension but were narrow in the $y$ (vertical) dimension.
The target centers were located at $z=-331.85$~cm.
Here, the $z$ axis is the direction of the incident proton beam, and
the origin of the right-handed coordinate system is centered at 
the upstream end of the SM12 yoke.

\subsection{Beam Monitors}	
\label{amon}
Beam intensity was measured using both an ion chamber and a secondary-emission 
monitor,~SEM, located upstream of the target.  
The fraction of beam striking the target was determined using an interaction 
monitor,~AMON, which was a scintillation-counter telescope 
perpendicular to the beam and  
viewed the target through a hole in the shielding cave.
The targeting fraction varied from 30\% to 40\% depending on 
the running conditions (see~\cite{Chen:1993} for details).

\subsection{Silicon Vertex Detector}
\label{ssd:chap2}
\begin{figure}[hbt]
\centerline{\psfig{figure=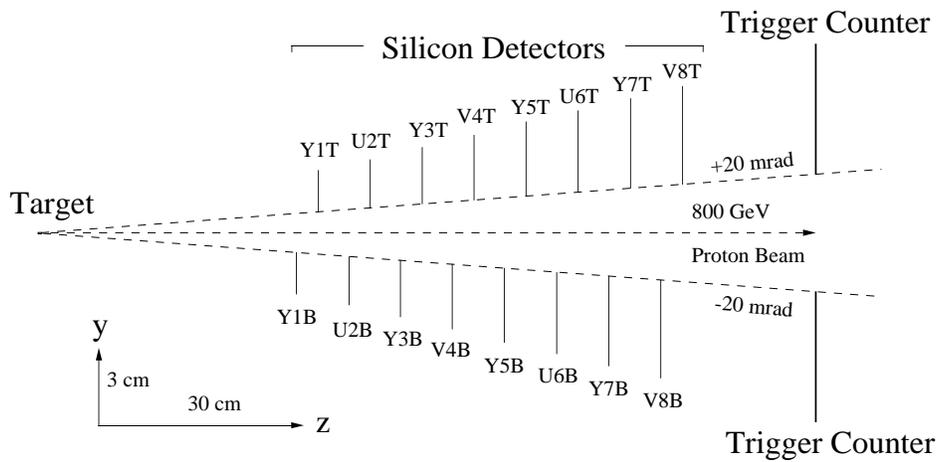,width=5in}}
\caption{Elevation view of E789 target region.\label{fig:ssdpic}}
\end{figure}

The SSD was located just downstream of the target and consisted of 
two arms, each containing eight planes of detectors 
(see Figure~\ref{fig:ssdpic}). 
Each 5-cm by 5-cm plane was a 300-$\mu$m-thick silicon-strip detector with
50-$\mu$m strip pitch.
The planes were arranged in two arms to cover vertical angles from  20 to
60~mr above and 
below the beam. 
Each plane had one of three orientations, Y, U, or V, 
with rotations about the $z$ axis of $0^{\circ}$, $+5^{\circ}$, or 
$-5^{\circ}$, respectively.  
The sequence of orientations in each arm was Y~U~Y~V~Y~U~Y~V,
proceeding downstream.  

A total of 8,544~strips were instrumented with amplifiers~\cite{Yarema}, 
discriminators~\cite{Turko}, and 
latches~\cite{CRs} having $\approx$30-ns effective time resolution.  
To minimize secondary interactions, thermal 
fluctuations, and radiation-induced detector degradation, the SSD containment 
volume was temperature-controlled with a $10^{\circ}$C helium fill.
Table~\ref{ssdtab} gives the configuration of the SSD planes.
One 1-mm-thick scintillator, of dimensions 5~cm$\,\times\,$5~cm, was
placed at the downstream end of each arm and was used
for triggering.

\subsection{Beam Dump}
\begin{figure}[hbt]
\centerline{\psfig{figure=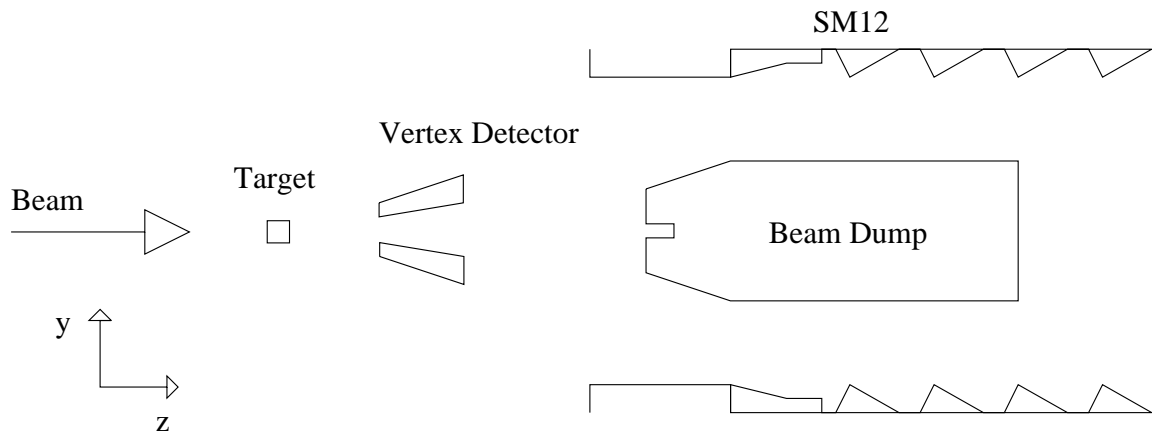,width=6in}}
\caption{Schematic diagram of E789 layout from target to beam dump (not to
scale).}
\label{fig:beamdump}
\end{figure}

A water-cooled copper beam dump was located inside the SM12 magnet 
(see Figure~\ref{fig:beamdump}).
It prevented noninteracting primary protons and secondary particles of 
low transverse momentum from entering the downstream spectrometer.  
The tapered dump and the baffles on the inside walls of SM12
defined the spectrometer aperture.

\subsection{Spectrometer Magnets}
The two dipole magnets, SM12 and SM3,
served to focus charged particles of momenta within a desired range
onto the downstream detectors.
The magnetic fields of both magnets were
carefully mapped with the Fermilab Ziptrack~\cite{Ziptrack}, 
and the resulting profiles of the
field were used in the data analysis.

SM12 was a 1200-ton, 14.5-m-long,
open-aperture dipole magnet.  
The horizontal aperture was tapered to provide a gradually decreasing 
magnetic field.
At an operating current of 
900~A, SM12 provided a vertical transverse impulse
($p_{t}$) of 1.6~GeV/$c$, optimal
for studying a \DO decaying into two charged particles.  

The second bending magnet, SM3, 
was located between tracking stations 1 and 2.
It was a 3.4-m-long, open-aperture magnet and
provided a 0.91~GeV/$c$ vertical 
$p_{t}$ impulse. 
It deflected charged particles in the opposite direction as SM12, 
focusing them onto the subsequent detectors. 
In combination with the drift chambers SM3 provided momentum analysis for 
charged particles. 

\subsection{Tracking Stations}
Three drift-chamber tracking stations were used to determine 
charged-particle trajectories through the spectrometer.
Each station consisted of three pairs of chambers, with 
the chambers in each pair offset by half a drift cell to resolve the
``left-right" tracking ambiguity.
Each pair was oriented in one of three views,
Y, U, and V, at $0^{\circ}$, $+14^{\circ}$,
and $-14^{\circ}$ with respect to the $y$~axis. 
The drift gas was a 50/50 mixture of argon and ethane, with 
a 0.7\% admixture of ethyl alcohol.  
Each sense wire was connected to its own time-to-digital 
converter to measure the drift time of the ionization electrons. 
At operating voltages around 2000~V, the drift velocities averaged 
about 50~$\mu$m/ns.
Hodoscope planes
at each tracking station provided fast coarse
tracking information used in the trigger. Stations 1 and 3 had both X and Y 
hodoscope planes (designated HY1, HX1, HY3, HX3), while station 2 had only
a Y plane (HY2).  Each 
hodoscope plane consisted of two half-planes of scintillation counters,
whose light was collected using lucite light guides glued to 
Hamamatsu~R329 photomultiplier tubes. 

\subsection{Calorimeter}
\label{cal:chp2}
\begin{figure}[hbt]
\centerline{\psfig{figure=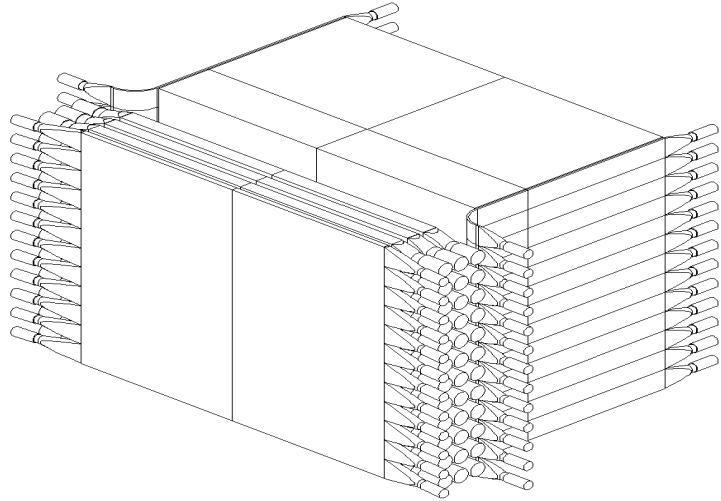,height=3in,width=4in,angle=90}}
\caption{Isometric view of E789 electromagnetic and hadronic calorimeters.}
\label{fig:cal}
\end{figure}

Sampling calorimeters were used to identify electrons and hadrons
(see Figure~\ref{fig:cal}).  
The electromagnetic section consisted of four lead/scintillator
layers, E1, E2, E3, and E4, with thickness of 2, 5, 5, and 6 radiation lengths, respectively. 
The total thickness of the electromagnetic section was 0.81 interaction length.
Each layer had separate
left ($x>0$) and right ($x<0$) sections which were divided into twelve modules
in $y$. Each of the resulting 96 modules  was
read out individually, with the analog signal from the scintillator
converted to a digital signal using an 8-bit quadratic ADC~\cite{KaplanADC}. 

The hadronic section consisted of two iron/scintillator layers, H1 and H2, 
of 2.14 and 5.84 interaction lengths respectively.   
The left and right segments of each were
divided into thirteen modules in $y$, giving a total of 52 modules. 
For details regarding the calibration of the calorimeters see \cite{Pripstein}.

\subsection{Muon Station}
\label{muonstation}
The Muon Station, located at the downstream end of the spectrometer, 
contained three planes of proportional-tube arrays and two planes of 
hodoscopes, interspersed with shielding.  
The proportional tubes, used in the Trigger Processor and 
for off-line muon identification, 
consisted of two planes of horizontal cells, PTY1 and PTY2, for 
$y$ determination and one of vertical cells, PTX, measuring the $x$
coordinate.  Each 
plane was made of a series of two-layer aluminum extrusions, each containing
fifteen 2.54~cm~$\times$~2.54~cm cells.
The layers were offset by half a cell width from each other.  
The cells were read out with latches, so no timing information was recorded.  
The gas mixture used was the same as in the drift chambers.  

The muon hodoscopes were used both for 
triggering and for particle identification.  There were two planes, HY4 and
HX4, providing $y$ and $x$ information respectively.  
The calorimeter and additional zinc, lead, and concrete 
shielding comprised 16 interaction lengths of material between 
station 3 and the muon station.  
In addition, concrete absorbers interspersed among the muon detectors
added approximately 5 more interaction lengths of shielding.  

\section{Data Acquisition}
The data-acquisition system~\cite{Crittenden}  was based on
the Nevis Laboratories Data Transport System~\cite{transport}.
Event information from the front-end crates was buffered into
multiport memory modules and supplied to the 
Trigger Processor~\cite{Schub:1996}
(described in Section~\ref{trigproc}) before readout to 
the VME-based archiving system, which recorded data on
four Exabyte 8200 tape drives.  The system was
capable of streaming approximately 1~MB/s to tape. 
Once per spill, scalers were read out to record
trigger rates and beam-intensity
information with and without system deadtime.

\section{Trigger}
\label{trigger}
E789 utilized a three-level trigger system.
Level-1 triggers based on hodoscope information
were OR'ed together to form the Trigger Fan In (TFI) signal.
Events satisfying TFI were latched and fed to the ``DC Logic" 
system, in which further logical requirements were imposed. 
In the DC Logic, information from slower spectrometer components, such as the 
calorimeter, could be included.
Events satisfying the DC Logic requirements produced the Trigger Generator 
Output (TGO) signal, which vetoed the fast latch reset, preserving hit 
information for readout to the Trigger Processor.
Finally, the Trigger Processor examined hit patterns in the wire chambers, 
hodoscopes, calorimeter, muon detectors, and silicon detectors to enhance 
the fraction of events in the desired decay channels that 
contained decay vertices downstream of the target. 
Events satisfying all three levels of 
trigger were written to tape.  In addition, at each 
trigger level, some events were prescaled and forced through to the next level. 

\subsection{TFI}
\begin{figure}[hbt]
\centerline{\psfig{figure=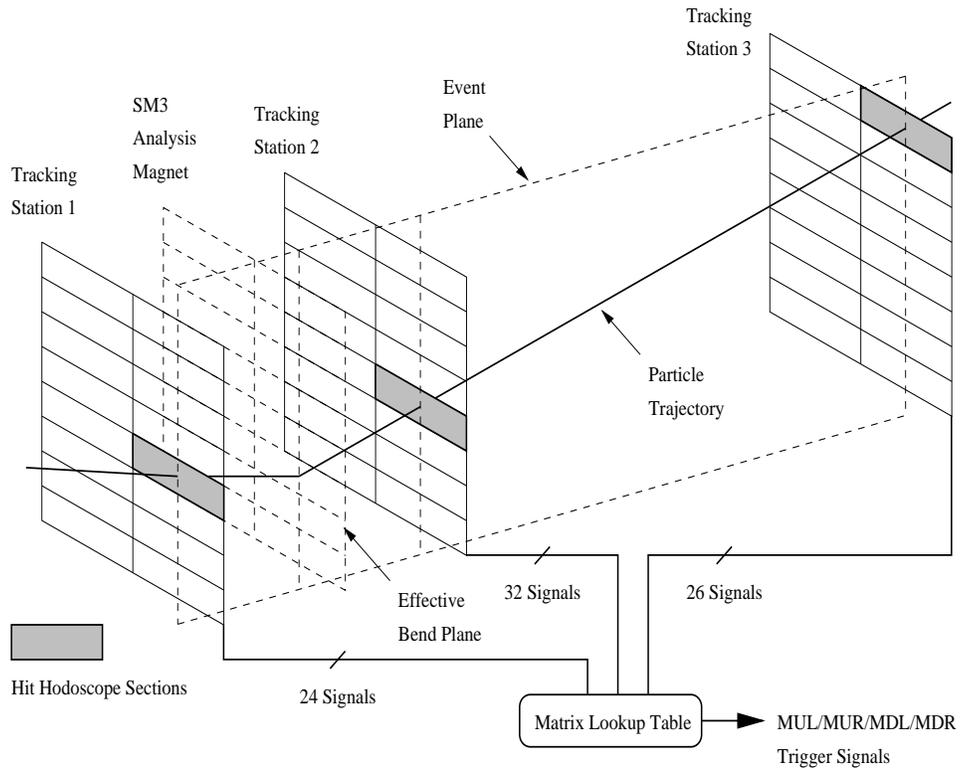,angle=-90,width=5in,height=4in}}
\caption{Conceptual sketch of hodoscope trigger matrix.}
\label{fig:matrix}
\end{figure}

The TFI had three main components, designed to 
trigger on pairs of charged particles from the target.
The main pair trigger, $\frac{2}{4} M$, required at least two triple 
hodoscope coincidences in HY1, HY2, and HY3, each corresponding to 
a different charged-particle trajectory from the target.
As shown in Figure~\ref{fig:matrix}, these 
``Trigger Matrix" coincidences were implemented as a look-up table, using fast 
ECL RAM, that provided four independent output signals corresponding to 
hodoscope roads to the left or right of the $z$ axis and passing above 
or below the beam dump.
At least two of these four outputs needed to fire to satisfy $\frac{2}{4} M$.

To allow sufficient redundancy for hodoscope and trigger
efficiencies to be determined off-line, additional triggers were implemented
using majority logic on combinations of the six hodoscope planes HX1, HY2,
HY3, HX3, HY4, and HX4. 
These were designated $\frac{n}{4} \mu LR$ and $\frac{n}{4} LR$.  
The notation $\frac{n}{4} \mu LR$ represents a 
logical AND of $\frac{n}{4} \mu L$ with $\frac{n}{4} \mu R$.  
The component $\frac{n}{4} \mu L$ required that at least $n$ of HX1, HY2, 
HX4 and HY4 had hits to the left of the $z$ axis.
Likewise $\frac{n}{4} \mu R$ required at least $n$ from the same group to 
have hits on the right side.
The trigger $\frac{n}{4} LR$ imposed a similar requirement except that 
planes HX1, HY2, HX3, and HY3 were used.  
In the run with SM12 current set to 1000~A, $n$ was set to 3.  
It was changed to 4 in the 900A run.

\subsection{DC Logic}
\label{dclogic}
The DC Logic, the second-level trigger, incorporated information from slower 
detectors whose use at the TFI stage would have imposed excessive deadtime.
To reject tracks missing the SSD planes,  
signals from the scintillators behind the SSD arms, $SU$ and $SD$, 
were required at this stage.
Veto signals ($\overline{NX1}$ and $\overline{NX3}$), formed by counting 
the number of hit counters in the hodoscopes HX1 and HX3, 
were used to veto high-multiplicity events.
The DC Logic also included particle-identification 
components based on the calorimeter and the muon station.
The various logic combinations were OR'ed together to form the TGO signal.

The DC Logic event-identification requirements caused differences in acceptance 
for various types of events.
The dimuon TGO component ($\mu^{+} \mu^{-}$) 
required that $2HX4$ and $2HY4$ be satisfied, that is, that two 
counters in each of the muon hodoscopes HX4 and HY4 register a hit.
The calorimeter provided a sum of analog signals from the dynodes 
of the E2 and E3 photomultiplier tubes which was sensitive to electrons, 
and another sum based on signals from H1, H2, E1 and E4 for hadrons.  
Each sum was discriminated with a low and a high threshold separately.  
The low threshold was set for detecting single particles and the high 
threshold for two-particle events. 
The discriminated signals 
$e$, $E$, $h$, and $H$ represented the low and the high thresholds 
for the electromagnetic and hadronic energy sums respectively.
Table~\ref{trigtab} lists the various logical combinations that together 
formed TGO.
As shown in Table~\ref{trigextab}, 
the DC Logic requirements reduced the trigger rate significantly 
relative to TFI.

\subsection{Trigger Processor}
\label{trigproc}
If an event satisfied one of the DC Logic triggers, the Trigger
Processor then
searched for tracks from the target using the drift-chamber information.
Only wire positions were used at this stage.
Track hypotheses were formed from hits 
in the Y drift chambers in stations 1, 2, and 3, 
masked by hodoscope and calorimeter or proportional-tube hits. 
These potential target tracks were then projected to the SSD, 
and used to identify SSD hits in the $y$-$z$ view for SSD trackfinding.
SSD tracks formed from the masked hits were subjected to the requirement 
that the impact-parameter be more than 51~$\mu$m from the center of the target, 
designed for finding tracks coming  
from $D^0$ decays occurring downstream of the target. 
The chosen tracks were then combined in up-down pairs to form vertices.
A cut of 0.10~cm was made on the location of the vertex in $z$ to further 
increase the likelihood that the event contained a downstream decay.  

The Trigger Processor reduced the trigger rate by about an 
order of magnitude below the TGO rate.
The Triggers After Processor (TAP) was dominated by the dihadron trigger.
During the ``Dedicated Dilepton" running period, the dihadron
trigger was prescaled by a factor of 32 and the proton intensity was increased 
to enhance the dilepton sensitivity.
Table~\ref{trigextab} gives the average rates per spill for protons on 
target, TFI, TGO, and TAP. 

\section{Data Sets}
\label{data}
Data was taken initially with SM12 set at 1000 A and two different target 
materials for studying the nuclear 
dependence of proton-induced charm production~\cite{Leitch:1994}.
Subsequently data was collected at 900 A because of improved acceptance for 
detecting charm decay at this setting. 
Three data sets are included in this analysis: 1000A-Au, 900A-Au, and 900A-Be.
Each set was processed separately and each yielded an independent normalization 
signal in the \dkpi\ mode.
(The latter part of the 900A-Au sample, for which the dihadron trigger was
prescaled as just described, is referred to as the Dedicated-Dilepton run, but 
it shared a common normalization signal with the rest of the 900A-Au sample.)
Table~\ref{datatab} gives the total number of protons on target 
for each sample, the number of $AMON\cdot\overline{SB}$ counts\footnote{
{\em AMON} is proportional to the number of interactions in the target
and is described in Section~\ref{amon}.  $\overline{SB}$ is {\em true}
when the system is able to accept data.
$\frac{AMON\cdot\overline{SB}}{AMON}$ is thus a measure of the 
{\em live time} of the data acquisition and was typically about
50\%.} (number of live-time-corrected counts in the targeting 
monitor), and the number of triggers recorded on tape.

\section{Normalization Approach}
To determine the branching ratios for \dee, \dmue, and \dmumu, we need
the total number of $D^0$s produced, as 
measured by the decay mode $D^{0}
\rightarrow K\pi$, as well as the detection efficiency for each decay
mode:

\begin{eqnarray}
\label{branch}
B(\dll) & 
= & \frac{N_{l^{+}l^{-}}
}{\epsilon_{l^{+}l^{-}}}\times \frac{\epsilon_{K\pi}
}{N_{K\pi}}B(D^{0}\rightarrow K\pi) 
\\
\vspace{0.3in} 
& \equiv & \epsilon\times\frac{N_{l^{+}l^{-}}
}{N_{K\pi\ }} \times B(D^{0}\rightarrow K\pi)\,.
\end{eqnarray}
Here 
$N_{l^{+}l^{-}}$ is the number of $\dll$ events seen, $N_{K\pi}$
is the number of \dkpi\ events, $\epsilon_{l^{+}l^{-}}$ is the
efficiency for observing a $\dll$ event, 
$\epsilon_{K\pi}$ is the efficiency for observing a \dkpi\ event, 
$B(D^{0}\rightarrow K\pi$) $=
(3.85 \pm 0.09)\%$ is the branching ratio for \dkpi\
decay~\cite{PDG:1996}, and 
$\epsilon \equiv \epsilon_{K\pi}/\epsilon_{l^{+}l^{-}}$ is the 
relative efficiency.
Using \dkpi\ as a normalization mode allowed partial 
cancellation of many common correction factors in the efficiency ratio.

\section{Event Reconstruction}
With a total nuclear inelastic cross section per nucleon of 17~mb 
for beryllium and an inclusive \DO production cross section of $
\approx 40~\mu$b at $\sqrt{s}=39$~GeV \cite{Kodama:1991},
a branching ratio $B(D^{0}\rightarrow K\pi)$ of $\approx4\%$ 
implies a search for one normalization event per ten thousand interactions. 
In this experiment, with an average
\DO decay distance of $\approx 3.4$~mm (corresponding to an average \DO momentum
of 56~GeV/$c$), precise reconstruction of the decay vertex in the SSD is a
powerful tool to separate \DO decays from background processes
that occur in the target.
The SSD allowed precise reconstruction of the decay distance 
and the impact parameter for each track. 
The impact parameter could be used to eliminate tracks 
from the target and thus reduce the background significantly.  

\subsection{Pass One}
\label{pass1}
In the first pass of data processing particle trajectories in both the
downstream spectrometer and the SSD were reconstructed from the raw data. 

Downstream track reconstruction began by finding hit clusters in 
the drift chambers.
Track segments formed in stations 2 and 3 were projected to station 1 for 
confirmation that the track came from the target and not the beam dump.  
An 18-plane (3 stations with 6 chambers each) least-squares fit 
was performed.   

The track momentum was determined from the bend angle through SM3.  
The momentum resolution of the spectrometer was found to be $\sigma /p$
= $1.58\times 10^{-4}p$, where $p$ is the momentum of a track and $\sigma$ 
is the statistical uncertainty.
The track was then traced back iteratively through SM12, through the SSD, 
to the target.   
On the first iteration the track was traced from SM3 to the $z$ 
location of the target center.
Candidate tracks falling within an aperture of $\pm 12.7$~cm from
the target center in $x$ and $y$ were kept. 
In subsequent iterations, the track parameters were adjusted so that the 
track traced back to the target center.

After downstream tracking, all track segments in the SSD were reconstructed.
In each of the SSD arms, the four Y planes were used first.
The preliminary $y$-$z$ tracks were then employed to define windows 
in the U and V planes for selecting hits that were used to form SSD tracks in 
the $x$-$z$ view.  
Those SSD tracks with enough hits in the Y, U and V views 
were fit to straight lines in three dimensions.
The resolution of the impact parameter in $y$ was determined to be 
$34~\mu$m.

For each event, each opposite-sign pair of downstream tracks was 
reconstructed as 
though they decayed from a single parent through each of the decay
modes \dkpi, \dmumu, \dmue, and \dee. 
For at least one of these modes, the resulting invariant mass 
was required to fall 
within a 500~MeV/$c^2$ window extending from 1.65~GeV/$c^2$ to 
2.15~GeV/$c^2$.
In addition, events were required to have at least one opposite-sign 
pair of SSD tracks that formed a vertex with $z$ location 
outside a $\pm 2.55$~mm window centered at the target.  
Since the resolution of the vertex in $z$ was about 0.7 mm, this 
requirement rejected a significant fraction of the dihadron events 
originated from the target but still retained about 50\% of 
the $D^{0}$ decays.
This pass provided a four-to-one data reduction from the raw data.  

\subsection{Pass Two}
\label{pass2}
\begin{figure}
\centerline{\psfig{figure=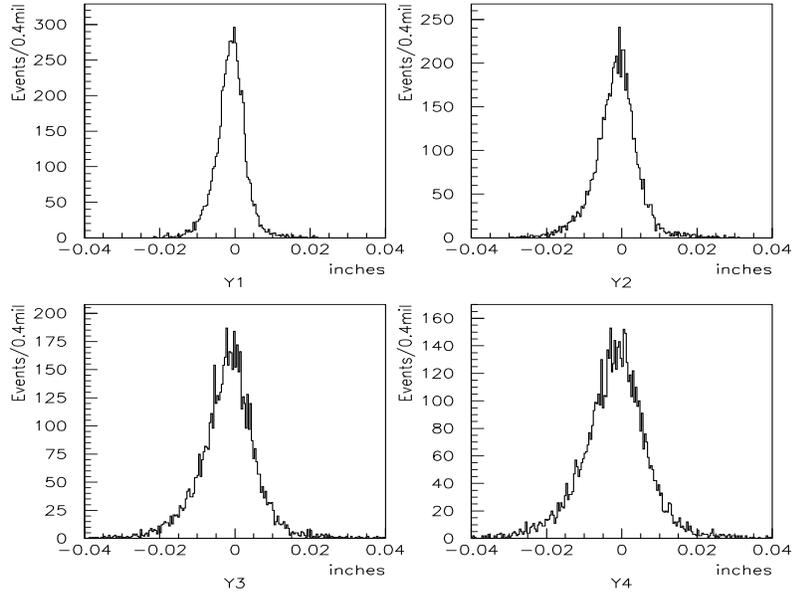,height=3.5in,width=4.5in}}
\caption{Distance between the downstream track and the associated SSD hit
for the Y planes in the lower SSD arm for events with only one SSD
track in either arm.}
\label{fig:yssdmatch}
\end{figure}

\begin{figure}
\centerline{\psfig{figure=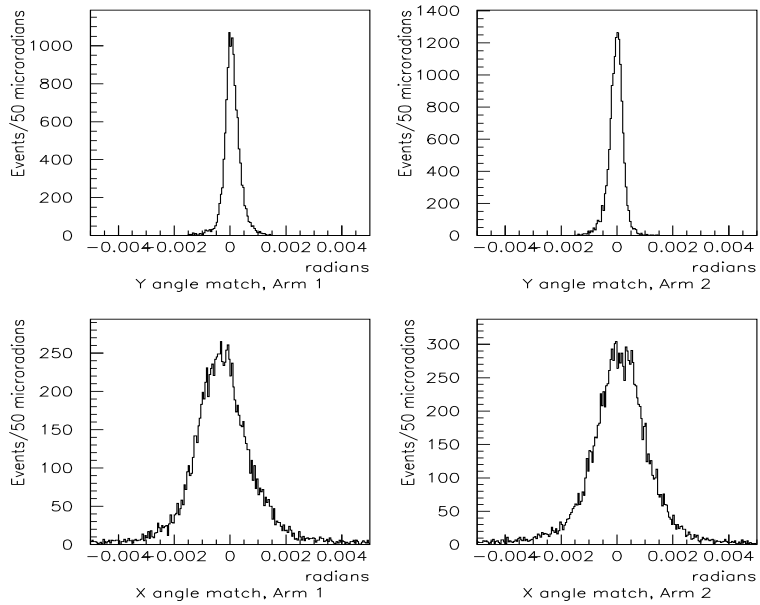,height=3.5in,width=4.5in}}
\caption{$y$- and $x$-angle matching between downstream track and SSD track
for both SSD arms for events with only one SSD track in either arm.}
\label{fig:yxangmatch}
\end{figure}

In this pass, to further reduce the number of unwanted SSD tracks, 
the $y$ hits, $y$-$z$ and $x$-$z$ angles of the SSD tracks were 
required to be within $\pm 0.2$cm, $\pm 1.5$~mr and $\pm 2.85$~mr 
respectively from the projections of the downstream tracks at the SSD.
Figures~\ref{fig:yssdmatch} and \ref{fig:yxangmatch} show 
the matching in $y$ and in 
track angles respectively, before the cuts, for events with 
only one SSD track in either arm.  

The matched SSD tracks, one from each arm, were combined to form pairs. 
The \chix\ was minimized for each pair 
by adjusting the vertex location and the track orientation.
The downstream tracks were then iteratively traced back to the decay vertex 
as determined by the SSDs to improve the resolution of the track angles.
The requirement on the invariant mass of the event
was tightened to a 200~MeV/$c^2$ window about the \DO mass
(1.864~GeV/$c^2$~\cite{PDG:1996}) for at least one of the modes.
This second pass provided another factor of five reduction of the data set.

\subsection{Pass Three}
\label{pass3}
The alignment of the SSD was refined in this pass.
The change in the alignment constants was insignificant.
At this point, 
some of the events still contained multiple vertices that  
resulted either from multiple pairs of downstream tracks or from 
multiple SSD tracks matching a single downstream track.
Events were then excluded if more than four SSD tracks matched either downstream
track or more than ten vertices were reconstructed. 
To select the proper vertex in the surviving events, 
the quality of each vertex was evaluated using nine parameters. 
The value of each parameter was converted to a probability, 
with the overall probability taken as the product of the nine probabilities. 
The nine parameters were:
\begin{itemize}
\item $\chi^{2}/{\rm degree\ of\ freedom}$ for reconstructing each SSD track. 
({\em 2 parameters})
\item $\chi^{2}/{\rm degree\ of\ freedom}$ for the SSD vertex-constrained fit.
({\em 1 parameter})
\item $y$-angle match between each downstream
track and its SSD track. ({\em 2 parameters})
\item $x$-angle match between each downstream track and its SSD
track. ({\em 2 parameters})
\item $\chi^{2}/{\rm degree\ of\ freedom}$ 
for the position difference between the projection of each downstream track and
the SSD hit at each SSD plane.  ({\em 2 parameters})
\end{itemize}
Events with only one SSD track in an arm 
were used to obtain the standard deviations 
of the distributions for the $x$ and $y$ angle matching,
as 0.95~mr and 0.25~mr respectively.  
Only the vertex with the highest overall probability, 
along with the associated SSD tracks and downstream tracks, was employed in
the subsequent analysis.

In the final stage of event selection, 
a fully reconstructed event consisted of one opposite-sign pair of 
SSD tracks that matched one pair of downstream tracks.
The most effective variable to optimize the \dkpi\ signal was the
impact parameter of each SSD track with respect to the
target center in $y$ before the vertex-constrained fit.
Incorrectly reconstructed events
often contained at least one track originating in the target
that was thus reconstructed with small impact parameter.
In addition, a cut was made on the lifetime~significance,  
defined as
the ratio of the $z$ location of the vertex to the average decay distance   
of the $D^0$ in the laboratory frame, $\gamma \beta c \tau$.
Tables~\ref{mumucuts},~\ref{eecuts} and \ref{muecuts} summarize 
the requirements on the 
impact~parameter and lifetime~significance for all data sets.

\section{Particle Identification}

\subsection{Electron and Hadron Identification}
Electrons and hadrons were identified using the calorimeter.
The identification procedure included two requirements: first, that energy
deposited in the calorimeter match a track, and second, that the
profile of the energy deposition in the calorimeter be consistent with
either a hadron or an electron.

For each event, the ADC counts of all calorimeter modules 
were converted to energy. 
The reconstructed particle trajectory was then projected through each
layer of the calorimeter.  At each layer, the energies deposited in the
module on the trajectory and in its nearest neighbors were summed. 
A correction was applied for attenuation 
of scintillation light in the $x$ (readout) direction.
If the track had no other track within two modules at each
longitudinal layer, it was considered isolated and its total energy
as well as the ``EM fraction" was recorded. 
The EM fraction is the amount of energy
deposited in the electromagnetic portion of the calorimeter divided by
the total deposited energy. 

The energy resolution of the calorimeter was derived from the ratio 
of the deposited energy in the calorimeter to the magnetically-measured 
momentum of isolated tracks ($E/p$).
\footnote{The $E/p$ distributions
were energy-dependent with various mean and $\sigma$.
These differences were taken into account in the analysis as described in
\cite{Pripstein}.}
For the hadronic part of the calorimeter, the resolution was 
$\sigma$/$E$ = -0.018 + 0.91/$\sqrt{E}$.
The energy resolution of the electromagnetic calorimeter was measured to be 
$\sigma$/$E$ = -0.04 + 0.79/$\sqrt{E}$.

For an isolated track, the particle associated with the track was
labeled as an electron if the EM fraction was $\geq 0.95$
and the $E/p$ of the track was within 2.58$\sigma$ from the mean of the 
$E/p$ distribution for that energy bin.
A separate study, using $J/\psi \rightarrow e^{+}e^{-}$
decays from data collected in an adjacent
running period, found this cut to be $\ge 96\%$ efficient~\cite{Chen:1993}.
For hadrons, the EM fraction was required to be less than 0.7 and the 
2.58$\sigma$ $E/p$ cut was used.

If two non-muon tracks shared at least one module (which happened quite 
rarely), they could often be identified using the
energy deposition in each section (EM and hadronic) of the calorimeter 
separately.
The procedure for identifying the overlapping tracks is described in 
\cite{Pripstein}.

\subsection{Muon Identification}
\label{muid}
Muon identification depended primarily on the muon station.  
A track was projected to the muon station
and each detector plane was checked to see if there were hits in 
momentum-dependent hit windows.  
The windows, 3$\sigma$ wide, were determined from fits to 
residual distributions with respect to the projected track.
For a track to be categorized as a 
muon candidate, hits matching the projected track were required in
both muon hodoscope planes and in at least two
proportional-tube planes. The
hodoscopes had a time resolution better than one 19-ns accelerator-RF bucket. 
Requiring them to
have hits dramatically reduced the number of out-of-time tracks.

In addition to the muon station,
the calorimeter was also employed for muon identification.
99.99\% of muons under 100~GeV/$c$ left less than 35\% of their energy
in the calorimeter.  If a track passed the muon
hit criteria and had $E/p$ greater than 35\%, it was tagged as
ambiguous but still counted as a potential muon.  
The most likely mechanism for muons to have high $E/p$
was for them to overlap with a non-muon track in the calorimeter.  
To remove fake dimuon events due to a non-muon overlapping with a muon,
an isolation requirement was applied. 
In this case, each track in the
reconstructed dimuon was required to have unique muon-hodoscope hits and 
no more than one proportional-tube plane could contribute a shared hit. 

\section{Acceptance and Efficiency}
\label{eff}
It is necessary to calculate
the ratio of (acceptance $\times$ efficiency) for each dilepton decay 
to that for the normalization decay.  
This relative acceptance depends on both the kinematics and particle types
in each decay mode. 
While track-reconstruction efficiencies cancel in the ratio, trigger and 
particle-identification efficiencies do not and are determined as described 
below.

\subsection{Monte Carlo Simulation}
The Monte Carlo (MC) program, incorporating trigger and particle-ID 
information, was used to generate events in each dilepton 
mode as well as the normalization mode. 
The MC used the same alignment constants and magnetic field maps
as the data analysis.  
Multiple scattering including non-Gaussian tails was also included in the 
simulation.
Detector efficiencies were included, and noise hits in the silicon detectors 
(extracted from data) 
were added to each generated event.

\begin{figure}[hbt]
\centerline{\psfig{figure=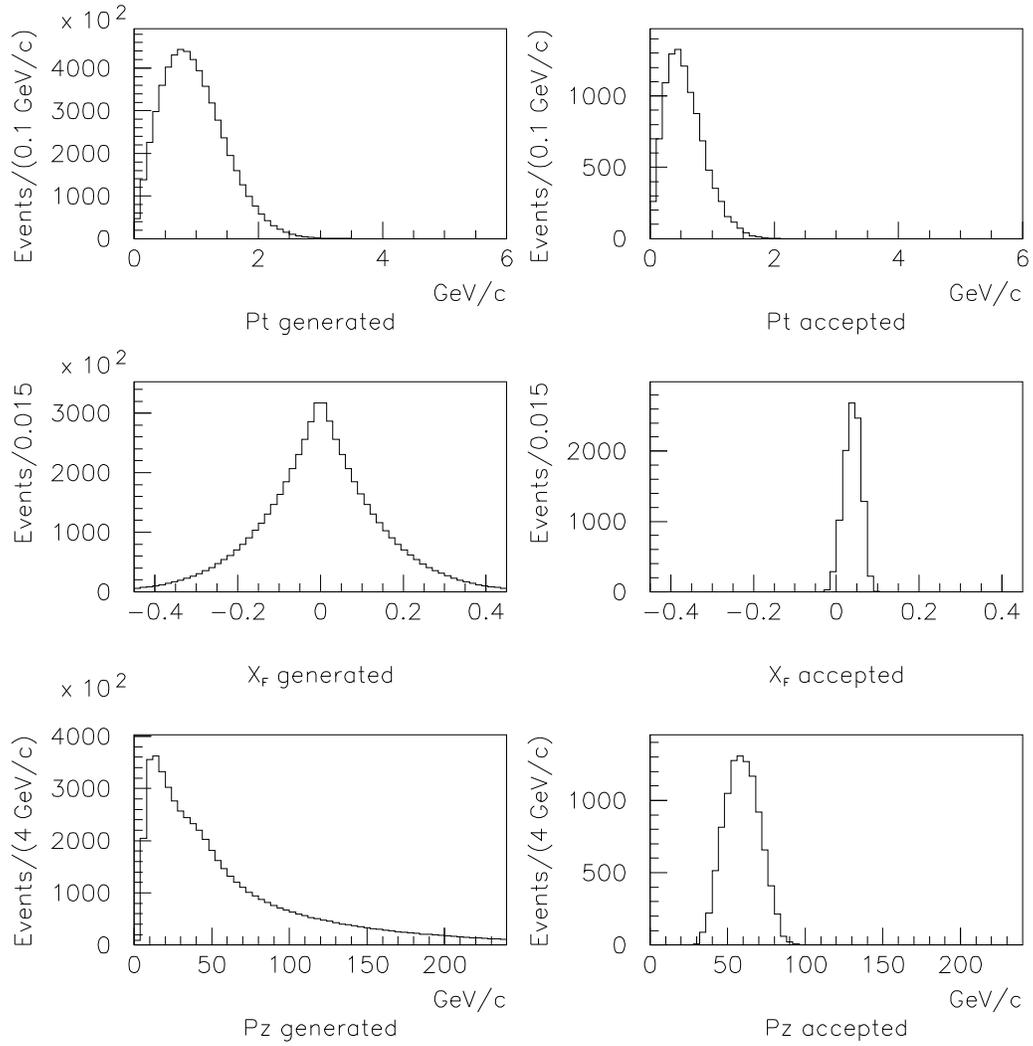,width=6in}}
\caption{Monte-Carlo generated (left) and accepted (right) distributions of 
$p_{t}$, $x_{F}$ and z component of momentum in the laboratory frame for 
$D^0$ with SM12 set at 900 A.} 
\label{fig:mcptxfpz}
\end{figure}

The production of $D^0$ by an 800-GeV proton beam was simulated
with a longitudinal fractional-momentum ($x_{F}$) distribution of the form  
$(1-|x_{F}|)^{6.9}$~\cite{Kodama:1991}.
The transverse-momentum ($p_t$) distribution was
characterized as $e^{-b  p_{t}^{2}}$ with 
$b=0.84\ ({\rm GeV}/c)^{-2}$~\cite{Kodama:1991}. 
Each two-body $D^{0}$ decay was generated with a uniform angular
distribution in the rest frame of the $D^{0}$. After boosting to the laboratory
frame, the decay products ($K\pi,~\mu\mu,~\mu e,\ {\rm or}\ ee$) 
were traced through the simulated geometry of the 
spectrometer. 
Kaon decay was also included in the Monte Carlo. 
Each event that passed the geometric restrictions was required to satisfy
the trigger as modeled for the decay of interest.
Figure~\ref{fig:mcptxfpz} 
shows the generated and accepted distributions in $p_t$, $x_{F}$, 
and momentum of the $D^{0}$ in the laboratory frame at 900~A.
With the kinematics of each decay properly modeled, the 
acceptances were calculated using 40,000 Monte Carlo events that
passed the geometric cuts for each mode.

\subsection{Dimuon Efficiency}
\label{dimueff}
Muons were identified primarily using the muon hodoscopes and proportional
tubes. 
The efficiency of each proportional-tube plane was
determined separately from data and then included in the Monte Carlo simulation.
Each muon selected for the efficiency study satisfied the muon identification 
requirements as described in Section~\ref{muid}.
This study determined
the average efficiencies of the proportional-tube planes to be 
95\%, 98\%, and 95\% for Y1, X, and Y2 respectively.

The only component of the dimuon trigger that was not included in 
the dihadron trigger\footnote{The dihadron-trigger efficiency was used in the
calculation of the relative
trigger efficiency between the \dmumu\ decay and the \dkpi\ decay.}
was the requirement of hits in at least two of the four quadrants
(upper-left, upper-right, lower-left, and lower-right) in both HX4 and HY4.
Determining the dimuon efficiency thus required an unbiased study of the 
muon-hodoscope efficiencies.  
A muon sample 
was chosen by selecting events that 
satisfied the calorimeter trigger and included at least one reconstructed 
muon, selected by requiring hits in at least two of the 
three proportional-tube  
planes and at least one muon hodoscope plane.
The window for finding the hodoscope hit was identical to 
the momentum-dependent window of the nearest proportional-tube plane.
The requirement of a muon hodoscope plane,
whose timing had single-bucket resolution, 
assured that the muon was indeed associated with the triggered event.  
The efficiency of each hodoscope plane was then 
determined by recording the fraction of events for which the hodoscope
that was not used in the muon-selection process 
fired.
The efficiencies of HX4 and HY4 were determined to be 95\% and 92\%
respectively, independent of muon momentum.
These efficiencies were then included in the MC simulation.

To avoid misidentification from overlapping tracks, an isolation criterion 
using the proportional tubes was applied.  The trigger
requirement already isolated the tracks such that the additional
proportional-tube isolation criterion reduced the efficiency by only 7\%
while reducing the background significantly.
The overall dimuon efficiency 
was 36\% at 900~A and 50\% at 1000~A.

\subsection{Dihadron Efficiency}

The efficiency of detecting the \dkpi~decay relative to that for the
dilepton modes is dominated by the efficiency of the $H$ trigger component, 
which required a significant amount of energy deposited in the 
calorimeter.

\begin{figure}[hbt]
\centerline{\psfig{figure=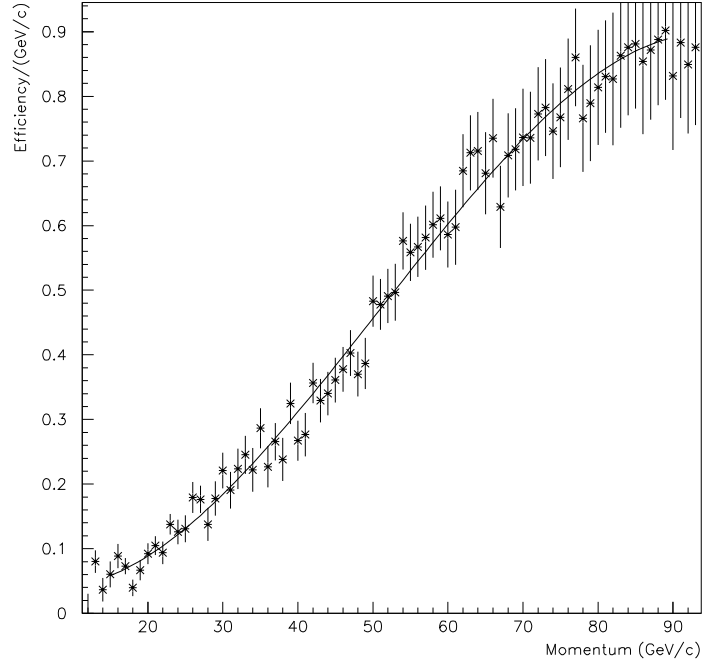,width=4in}}
\caption{Trigger efficiency for dihadron events as a function of pair momentum. 
The solid curve is the
parametrization used in the Monte Carlo program.} 
\label{fig:hhturnon}
\end{figure}

An unbiased sample of events passing the TFI trigger was employed 
for studying the $H$ efficiency.
Each event in this ``prescaled-TFI" sample had two hadron tracks with a 
reconstructed $K\pi$ invariant mass in a
500-MeV/$c^2$ window about the \DO mass.
The momentum range of the selected events was similar to  
that of the accepted \dkpi\ events. 
The fraction of the prescaled-TFI events
that fired the dihadron trigger was then plotted as a function of the total
momentum in 2~GeV/$c$ bins.
Figure~\ref{fig:hhturnon} shows the efficiency of the dihadron trigger 
as a function of momentum and the fit thereto
by a third-order polynomial. This efficiency curve was then
input to the Monte Carlo.
The average dihadron trigger efficiency for \dkpi\ events that 
passed the geometric acceptance of the MC was 55\% at 900~A and 58\% at 1000~A.

The energy deposition of hadrons in each section of the calorimeter 
was also included in the MC simulation.   
To enhance the certainty of hadron identification, the EM fraction of the MC  
events was required to be less than 70\%.  
This cut accepted over 92\% of hadrons.  
Furthermore, the $E/p$ of the particle was required to fall 
within $\pm$2.58$\sigma$ of the mean.

Kaon decay before Station 4 could cause
the event to be misidentified or could cause the
reconstructed invariant mass to drop out of the \DO mass window.
About 20\% of \dkpi\ events were lost due to kaon decay.

\subsection{Dielectron Efficiency}
\label{EEeff}
The trigger efficiency for the decay \dee\ relative to \dkpi\ was
dominated by the $E$ trigger component.
As discussed in section~\ref{dclogic},
$E$ was used for finding dielectron events while $e$ was used 
for single-electron events.
The same prescaled-TFI sample used for the
dihadron-trigger-efficiency study was used to determine the 
efficiency of the $E$ trigger.
The energy-sum signal of layers E2 and E3, E$_{sum}$, was 
digitized by an ADC for off-line study.
An efficiency curve as a function of the E$_{sum}$-ADC count was determined 
and then included in the Monte Carlo.  

\begin{figure}[hbt]
\centerline{\psfig{figure=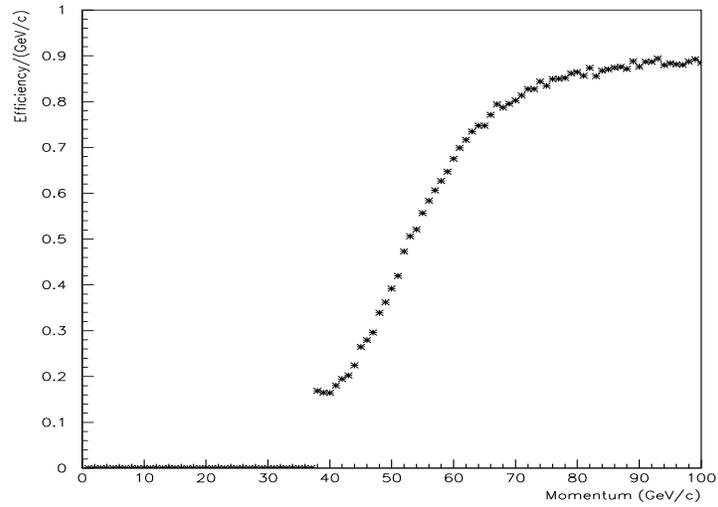,height=3.0in,width=4in}}
\caption{Efficiency of dielectron trigger as a function of Monte-Carlo 
generated \DO momentum. This efficiency was used for both 900A and 1000A runs.} 
\label{fig:eeturnon}
\end{figure}

Data events with an EM fraction greater than 0.95 and 
with tracks isolated in the calorimeter were used to determine
the energy deposited in E2 and E3 as a function of momentum.
When an electron was generated from the \dee\ decay, the energy that the 
electron deposited in the E2 and E3 calorimeter layers was generated  
based on the momentum-dependent (E2+E3)/$p$ distribution. 
Once the E2+E3 energy stored was determined, an ADC count was 
generated according to the E$_{sum}$-to-ADC curve.  
Figure~\ref{fig:eeturnon} shows the dielectron
efficiency as a function of the Monte-Carlo generated \DO momentum.  
The final trigger efficiency for \dee\ events that passed the Monte 
Carlo geometric cuts 
was determined to be 60\% for both 900A and 1000A runs.  

\subsection{$\mu$e Efficiency}
\label{MUEeff}
\begin{figure}[hbt]
\centerline{\psfig{figure=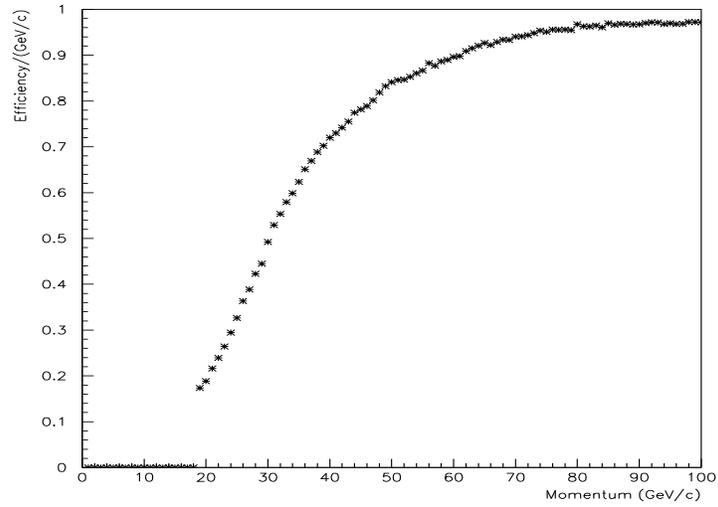,height=3.0in,width=4in}}
\caption{Efficiency of single-electron trigger as a function of 
Monte-Carlo generated \DO momentum. This efficiency was used for analyzing the 
900A and 1000A data.}
\label{fig:elowturnon}
\end{figure}

To find the efficiency for \dmue\ relative to \dkpi, we used the
techniques and tools developed for the \dmumu\ and \dee\ efficiency
analyses.  
To adapt to a single-muon event, we used a simple 2-hodoscope
requirement for the trigger, and demanded 2 out of 3 proportional-tube planes
to have hits.
The single electron was treated in a similar manner as the dielectron, 
the only difference being that the low-energy 
threshold (see Section \ref{EEeff})
required a different trigger-efficiency-to-ADC mapping.  As in the
dielectron case, the generated electron was required to
pass the geometric cuts of the Monte Carlo.  
Figure~\ref{fig:elowturnon} shows the turn-on curve for the single-electron 
trigger as a function of the Monte-Carlo generated \DO momentum.  
The resulting single-electron efficiency was
48\%, and the single-muon efficiency was 74\%, yielding a combined efficiency 
of 36\% for the 900A and 1000A runs. 

\section{Results}
\label{answer}
\subsection{Normalization}
\label{normalization}
\begin{figure}[hbt]
\centerline{\psfig{figure=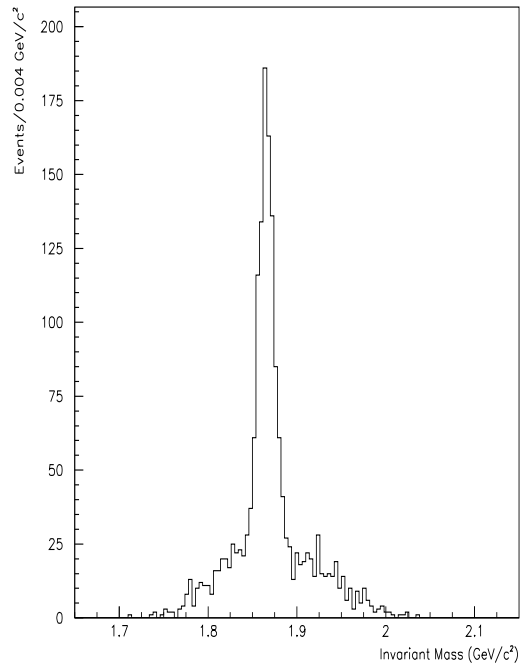,height=4in,width=3in}}
\caption{Reconstructed $K\pi$ invariant-mass distribution 
of Monte-Carlo \dkpi\ events at 900~A, with entries 
both for correct and for incorrect particle assignments.} 
\label{fig:mcmass}
\end{figure}

An event was labeled as a \dkpi\ candidate if it was reconstructed as a
dihadron event, satisfied the dihadron trigger, and passed the impact-parameter 
and proper-lifetime requirements that were applied to the dilepton decay.  
There was no mechanism to distinguish kaons from pions.
However, by Monte Carlo simulation we determined that the 
invariant-mass distribution of \dkpi\ events with incorrect particle
assignments was much
wider than that with the correct assignments, 7.1 times as wide for the
900A data sets and 5.4 times as wide for the 1000A set (see
Figure~\ref{fig:mcmass}).  
For each event, the invariant mass was thus computed once as $K^-\pi^+$ and once
as $\pi^-K^+$. 
The invariant-mass distribution was then fit to 
a quadratic polynomial for the background and a double Gaussian in the signal
region.
The standard deviation and normalization were allowed to vary for
the narrow Gaussian, but the width and relative height of the wide Gaussian
were constrained to the Monte Carlo values.  
This condition ensured that the numbers of events under the two Gaussian
distributions be identical.
In addition, both Gaussians were required to have the same mean mass.  
The fit was performed using 
PAW~\cite{Paw:2.05}.  The covariance matrix
for the width and normalization of the narrow Gaussian was then used
to find the absolute error associated with the number of reconstructed
\dkpi\ decays.  
Tables~\ref{mumunorm}, \ref{eenorm}, and \ref{muenorm} give the mean mass,  
the mass resolution and the total number of \dkpi\ events without any mass cut 
for each data set.
 
\subsection{Signals}
\begin{figure}[hbt]
\centerline{\psfig{figure=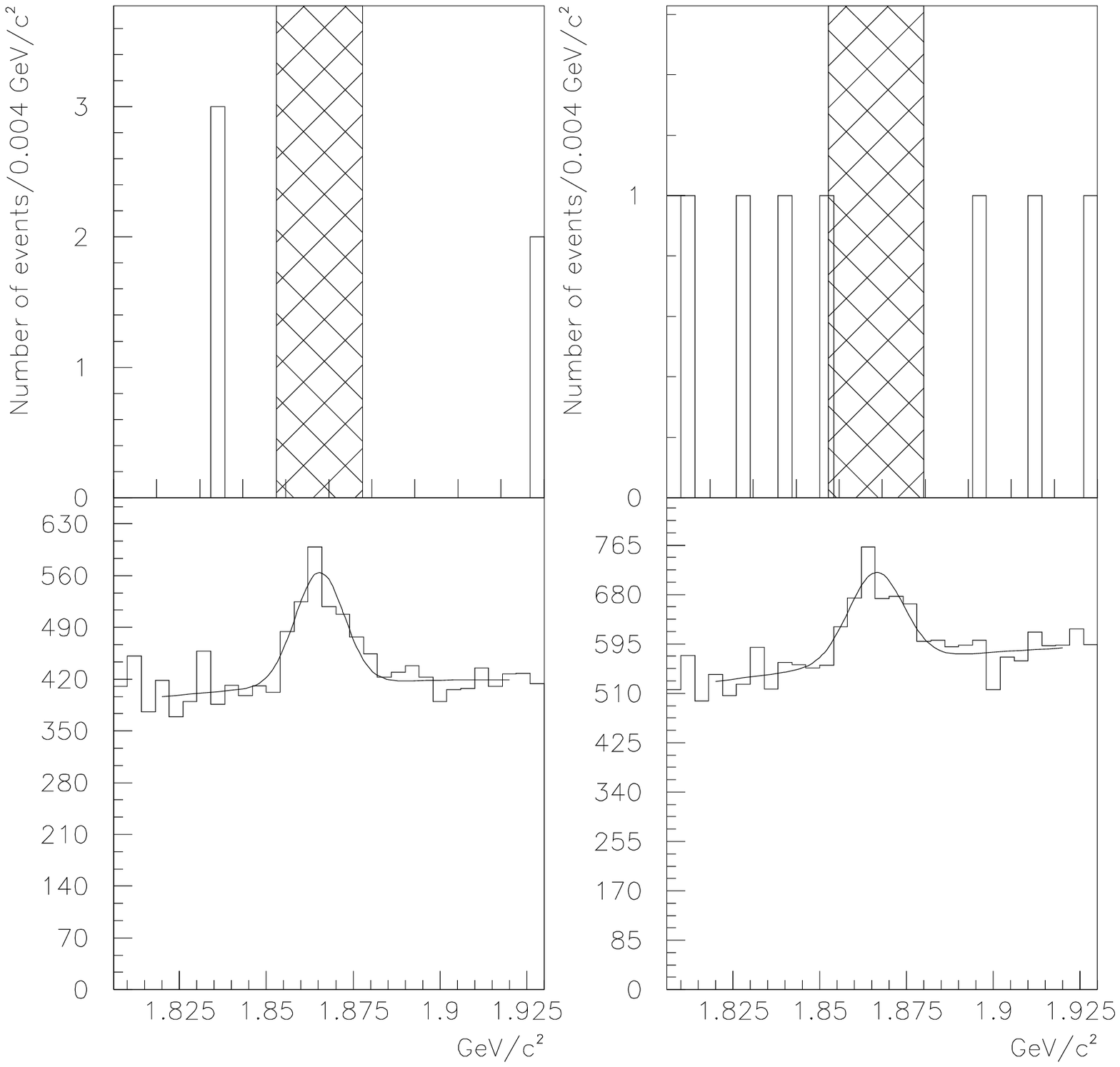,width=6in}}
\caption{Invariant mass distributions for \dmumu\ and its associated
\dkpi\ distribution for the 900A-Au and dedicated-dilepton data sets.
The cross-hatched area marks the signal region.}
\label{fig:au900mumu}
\end{figure}

\begin{figure}[hbt]
\centerline{\psfig{figure=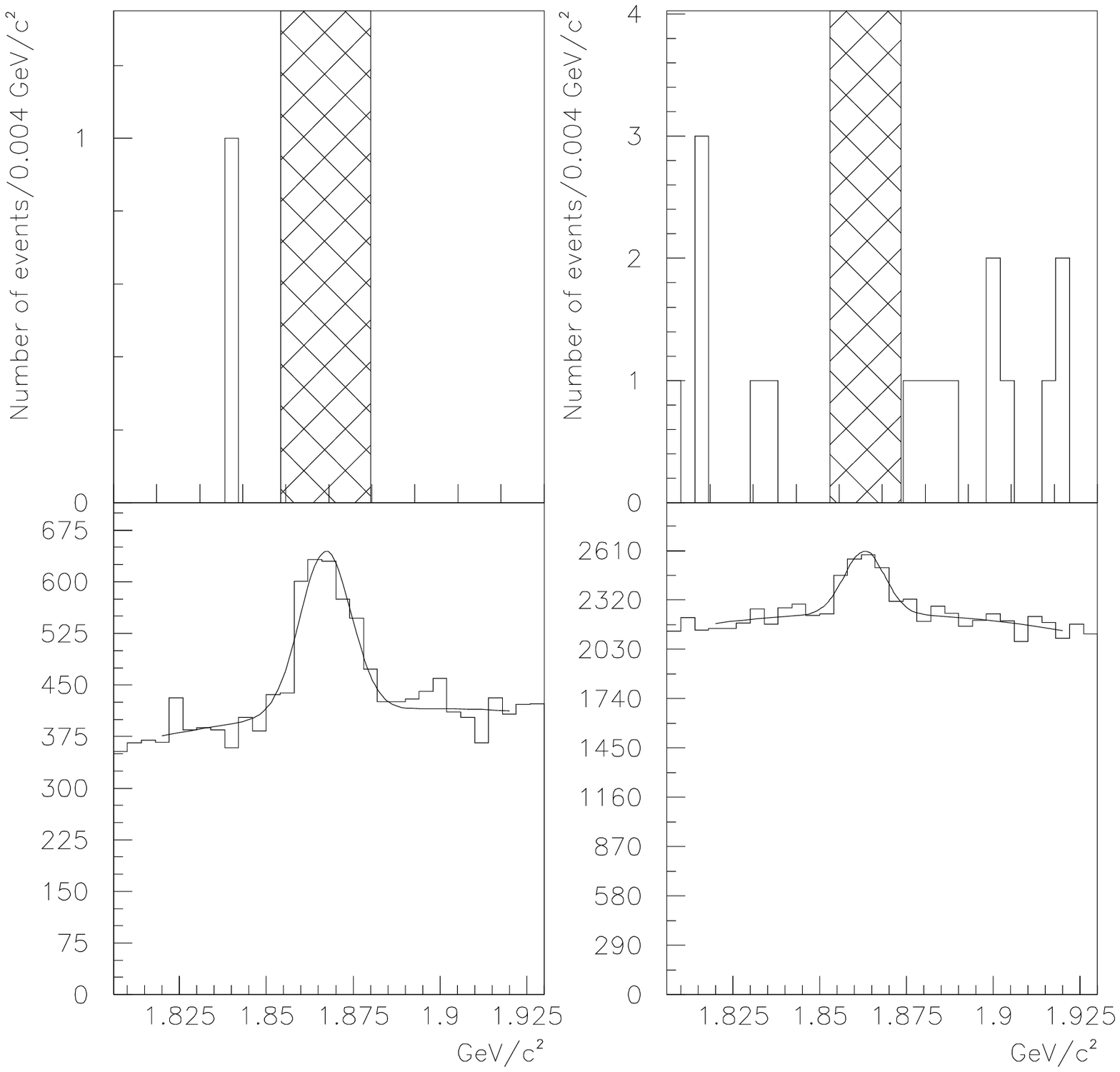,width=6in}}
\caption{Invariant mass distribution for \dmumu\ and its associated
\dkpi\ distribution for the 1000A-Au and 900A-Be data sets.
The cross-hatched area marks the signal region.}  
\label{fig:be900mumu}
\end{figure}

\begin{figure}[hbt]
\centerline{\psfig{figure=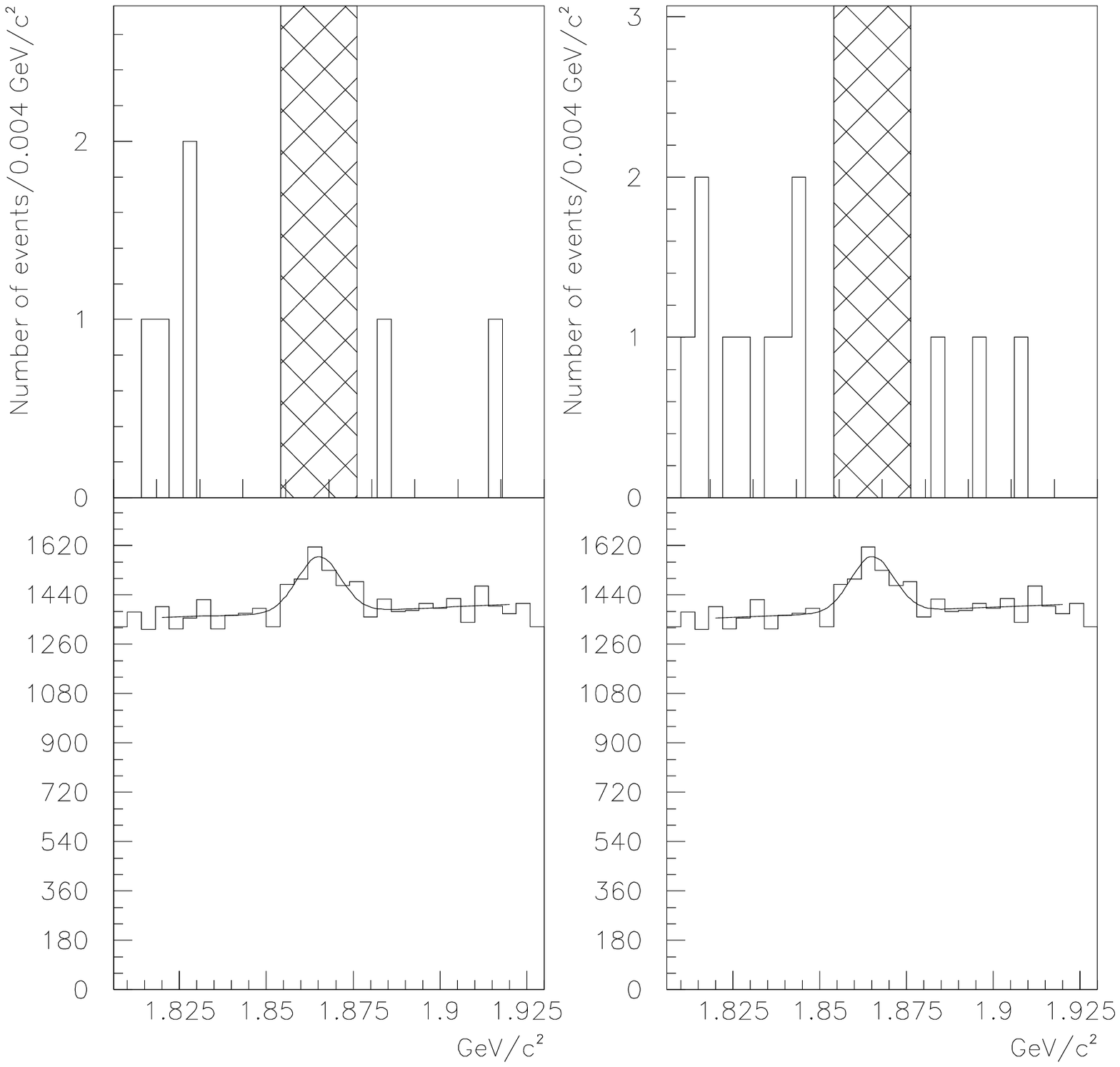,width=6in}}
\caption{Invariant mass distribution for \dee\ and its associated
\dkpi\ distribution for the 900A-Au and dedicated-dilepton data sets.
The cross-hatched area marks the signal region.}  
\label{fig:au900ee}
\end{figure}

\begin{figure}[hbt]
\centerline{\psfig{figure=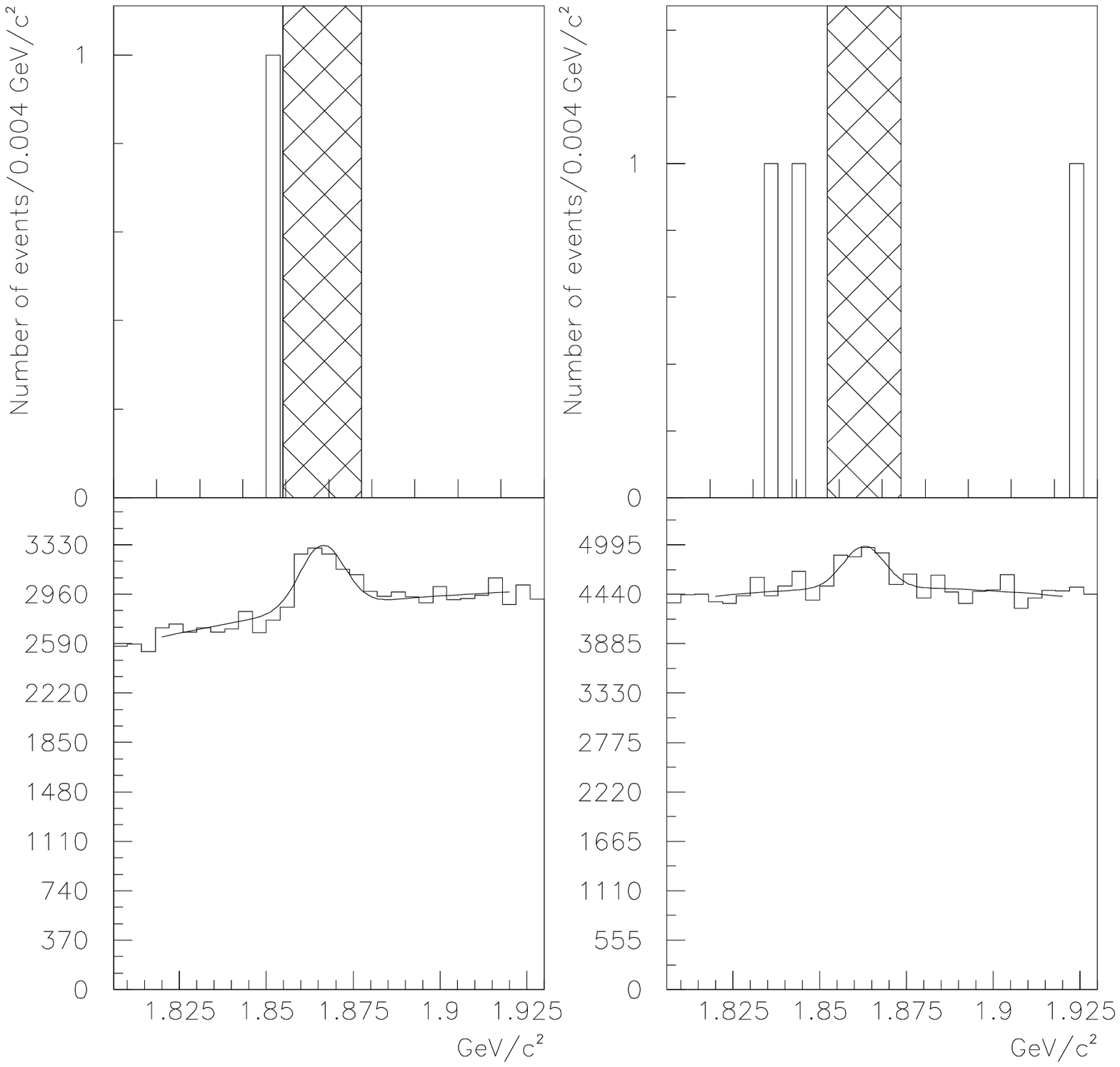,width=6in}}
\caption{Invariant-mass distribution for \dee\ and its
associated \dkpi\ distribution for the 1000A-Au and 900A-Be data sets.  
The cross-hatched area marks the signal region.}
\label{fig:be900ee}
\end{figure}

\begin{figure}[hbt]
\centerline{\psfig{figure=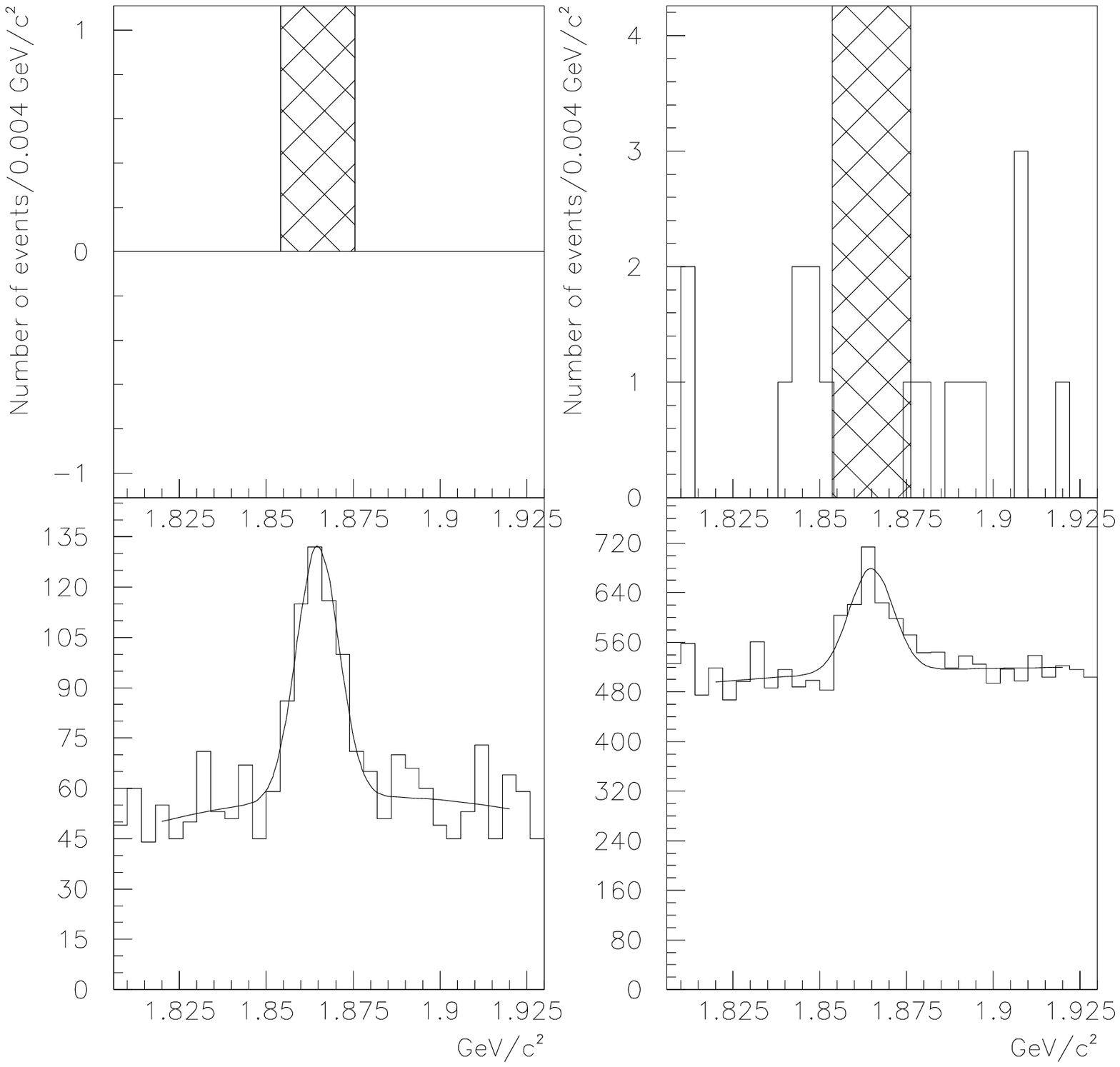,width=6in}}
\caption{Invariant mass distribution for \dmue\ and its
associated \dkpi\ distribution for the 900A-Au and dedicated-dilepton 
data sets.  
The cross-hatched area marks the signal regions.}
\label{fig:au900mue}
\end{figure}

\begin{figure}[hbt]
\centerline{\psfig{figure=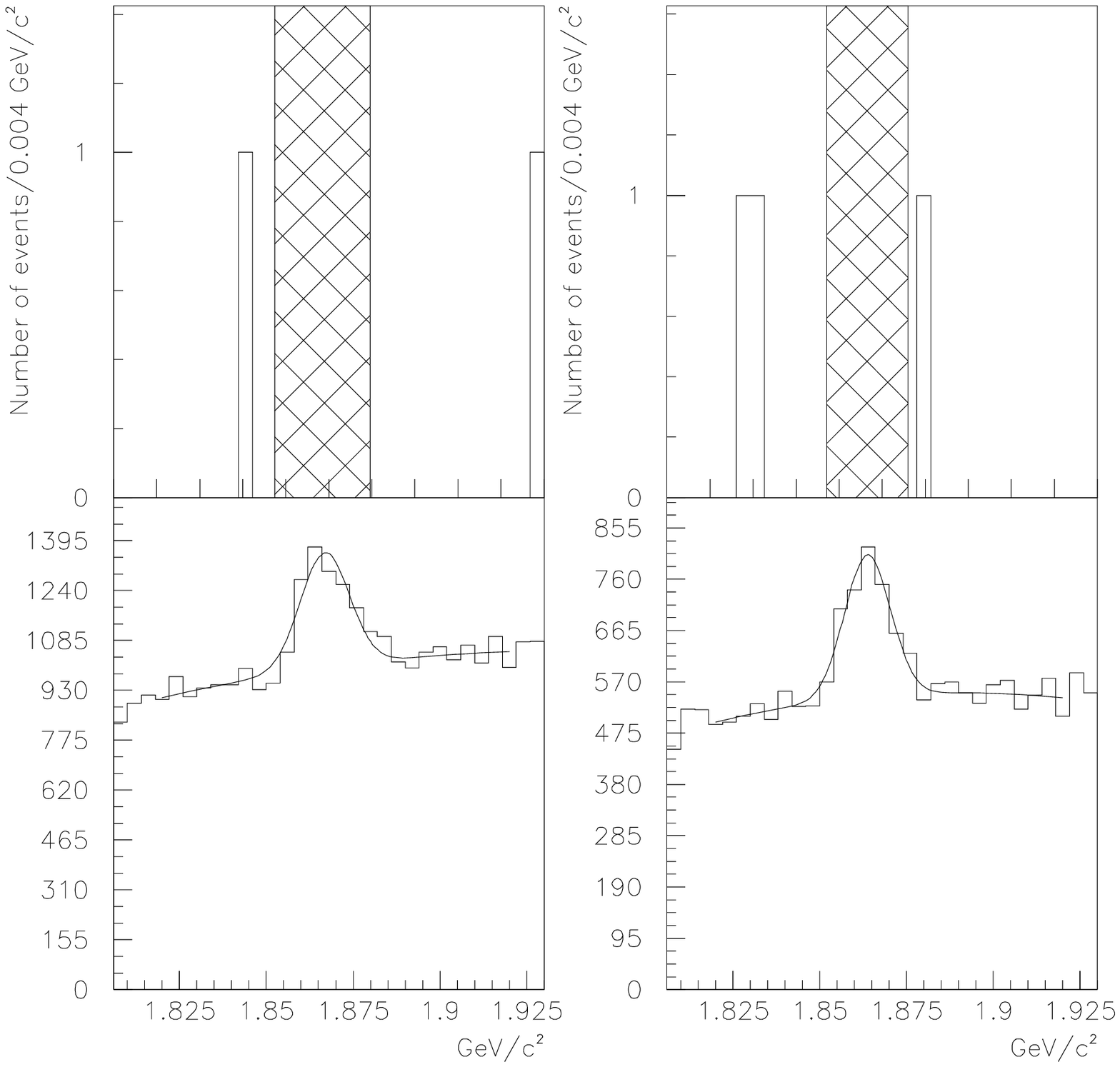,width=6in}}
\caption{Invariant mass distribution for \dmue\ and its associated
\dkpi\ distribution for the 1000A-Au and 900A-Be data sets.  
The cross-hatched area marks the signal region.}  
\label{fig:be900mue}
\end{figure}

Each data set listed in Table~\ref{datatab} was analyzed independently.  
When the impact-parameter and lifetime-significance cuts (see Tables~\ref
{mumucuts},~\ref{eecuts} and \ref{muecuts}) were applied to the dilepton 
data sample, 
as shown in Figures~\ref{fig:au900mumu}-\ref{fig:be900mue}, no 
event was found in 
the signal region, defined as the interval in dilepton invariant mass within 
which the signal events were counted.
This interval was $\pm1.96$$\sigma$ wide and centered at the mean of the 
$K\pi$ invariant mass of the corresponding normalization sample.
Since the mass resolution depended on the final-state particles, a different 
$\sigma$ was used for each dilepton decay as computed by MC and tabulated 
in Table~\ref{relwidth}.

\subsection{Summary of Acceptances and Efficiencies}

Tables~\ref{tabdkpieff} through~\ref{tabdmueeff} give the
various contributions to the acceptance\,$\times$\,efficiency 
for the normalization mode and each signal mode, together with their errors
as estimated in Section~\ref{systematic} below.

\subsection{Branching Ratios}
\label{upperlimit}

For each dilepton mode the 900A-Au, 900A-Au-Dedicated-Dilepton, 900A-Be, 
and 1000A-Au data sets were combined to give the branching ratio 
\begin{eqnarray}
\label{branch1}
B(\dll)  = \frac{ \sum_{i} S_{i} 
}{ \sum_{i} N_{i} \epsilon_{i} } \times B({ D^{0} \rightarrow
K\pi })\,,
\end{eqnarray}
where
$i$ runs over the three data sets, 
$S_{i}$ is the number of counts in the signal region in the 
dilepton-invariant-mass distribution,
$\epsilon_{i}$ is the efficiency for the dilepton mode given in 
Tables~\ref{tabdmumueff},~\ref{tabdeeeff} and \ref{tabdmueeff} 
relative to that for the
normalization mode in Table~\ref{tabdkpieff}, and 
$N_{i}$ is the number of observed $K\pi$ events in the signal region. 

With no signal observed, the upper limit on the branching ratio was 
determined using the Monte Carlo method. 
A series of branching ratios were calculated according to 
Equation~(\ref{branch}). 
For each calculation 
the expected number of counts in the signal region, $S_{i}$, 
was determined from Poisson statistics.  
Specifically, $S$ was distributed as 
\begin{eqnarray}
\frac{e^{-S}S^{M}}{M!}
\end{eqnarray}
with $M$ being the actual number of signal counts seen in the data.
The quantities $N_{i}, \epsilon_{i}$, and $B(D^{0}\rightarrow K\pi)$ were each
treated as a Gaussian distribution with the same mean and error used in 
Equation~(\ref{branch1}).   
A minimum of $10^{6}$ events were generated for each calculation.   
From the distribution of the calculated branching ratios, an upper limit 
on the branching ratio at the 90\%
confidence level\footnote{That is, 90\% of the generated branching ratios were  
less than this value.} was established.  
By this method, upper limits on
the branching ratios, at the 90\% confidence level, of
$1.56\times 10^{-5}$ for \dmumu, $8.19\times 10^{-6}$
for \dee\ and $1.72\times 10^{-5}$ for \dmue\ were obtained.

We have also employed the method of 
Cousins and Feldman~\cite{Cousins-Feldman},
as advocated by the Particle Data Group~\cite{PDG:1998}, for the case  
in which no signal and no background are observed.
In this approach, the upper limits on the branching ratio at the 90\% 
confidence level, corresponding to 2.44 events, 
are $1.65\times 10^{-5}$ for \dmumu, $8.69\times 10^{-6}$
for \dee\ and $1.82\times 10^{-5}$ for \dmue\ decay.
These results are about 10\% worse than those found by the Monte Carlo 
technique.
Table~\ref{results} summarizes the single-event sensitivity of our 
experiment and the upper limits for the $\dll$ decays as 
determined with the Monte Carlo approach and the Cousins-Feldman method.
It should be noted that the uncertainties in $N_{i},~\epsilon_{i}$, 
and $B(D^{0}\rightarrow K\pi)$ are not taken into account in the 
Cousins-Feldman method; hence, we favor the Monte Carlo method for 
determining upper limits.
Our \dee\ and \dmue\ limits are comparable to those set by 
CLEO~\cite{Cleo:1996}, whereas the \dmumu\ result is about a factor of 
four worse than those of 
BEATRICE~\cite{Beatrice:1997} and FNAL E771~\cite{E771:1996}.

\subsection{Systematic Errors}
\label{systematic}
Various systematic errors could affect the results in the branching-ratio
determination.
We follow Reference~\cite{Cousins:1992} in discussing the
effect of systematic errors on an upper-limit calculation.  

The uncertainties associated with the $p_{t}$
and $x_{F}$ distributions used in the MC simulation 
contribute to the error in the relative efficiency.  
This effect was studied by varying each parameter of the $p_{t}$ and $x_{F}$
parametrizations by $\pm1$$\sigma$.
The worst case was the variation in $x_{F}$, resulting in a shift in 
the absolute efficiency of $<17\%$.  
However, in each case the efficiency of the dilepton
mode relative to the normalization mode varied much less, only $\approx 2\%$.
The temporal variation of the dimuon, dihadron, dielectron, and $\mu e$-trigger efficiencies was investigated by subdividing the samples into 
independent data sets.  
As shown in Tables~\ref{tabdkpieff}-\ref{tabdmueeff}, the variations are small.
The impact-parameter and lifetime-significance cuts, varied by $\pm$1$\sigma$,  
did not have any significant effect on the relative efficiencies.

\section{Conclusions}

Three rare or forbidden decays, \dmumu, \dee, and
\dmue, have been searched for, and no evidence has been found for any of 
these decays.
New upper limits on the branching ratio at the 90\% confidence level 
are $1.56\times 10^{-5}$ 
for \dmumu, $8.19\times 10^{-6}$ for \dee\ and $1.72\times 10^{-5}$ for \dmue\
decay. For comparison, the best published limits on these decays are $4.1\times
10^{-6}$ for \dmumu\ ~\cite{Beatrice:1997,E771:1996},
 $1.3\times 10^{-5}$ for \dee\ ~\cite{Cleo:1996}, 
and $1.9\times 10^{-5}$ for
\dmue\ ~\cite{Cleo:1996}. 
Our limits for \dee and \dmue\ are the best to date. 
These limits, however, are still many orders of magnitude from the
levels at which these processes might be expected to occur.

\section{Acknowledgements}

We thank the staffs of Fermilab and the Los Alamos and Lawrence Berkeley
National Laboratories for their support. 
This work was supported by the Director, Office of Science, Office of High 
Energy and Nuclear Physics, of the U.S. Department of Energy under Contract No. 
DE-AC03-76SF00098, the National Science Foundation, and the National
Science Council of the Republic of China. 

\pagebreak

\pagebreak

\pagebreak

\pagebreak

\pagebreak
 
\pagebreak
 
\pagebreak
 
\pagebreak

\begin{table}[hbt]
\begin{center}
\caption{Dimensions of targets.\label{target_table}}
\begin{tabular}{ccc}
	&$900$A	&	$1000$A \\	
\hline
Length along $z$ (mm)	& 1.8	& 0.8\\
Height along $y$ ($\mu$m)	& 160	& 110	\\
\end{tabular}
\end{center}
\end{table}

\begin{table}[hbt]
\begin{center}
\caption{Configuration of Silicon Vertex Detector.\label{ssdtab}}
\begin{tabular}{ccccccc}
Plane &  & $z$ Position & $y$ Position &  &  & Number\\
No. & \raisebox{1.5ex}[0pt]{Name} & (cm) & (cm) & \raisebox{1.5ex}[0pt]{View} 
& \raisebox{1.5ex}[0pt]{Arm} & of Strips\\
\hline
1 & Y1B & -294.54 & -2.125 & Y & Lower & 316 \\
2 & Y1T & -291.36 & 0.949 & Y & Upper & 316 \\
3 & U2B & -286.92 & -2.300 & U & Lower & 372 \\
4 & U2T & -283.74 & 1.066 & U & Upper & 372 \\
5 & Y3B & -279.30 & -2.758 & Y & Lower & 436 \\
6 & Y3T & -276.12 & 1.548 & Y & Upper & 436 \\
7 & V4B & -271.68 & -2.865 & V & Lower & 500 \\
8 & V4T & -268.50 & 1.721 & V & Upper & 500 \\
9 & Y5B & -264.07 & -3.364 & Y & Lower & 572 \\
10 & Y5T & -260.88 & 2.217 & Y & Upper & 572 \\
11 & U6B & -256.44 & -3.566 & U & Lower & 628 \\
12 & U6T & -253.26 & 2.289 & U & Upper & 628 \\
13 & Y7B & -248.82 & -4.018 & Y & Lower & 692 \\
14 & Y7T & -245.64 & 2.805 & Y & Upper & 692 \\
15 & V8B & -241.20 & -4.154 & V & Lower & 756 \\
16 & V8T & -238.02 & 2.925 & V & Upper & 756 \\
\end{tabular}
\end{center}
\end{table}

\begin{table}[hbt]
\begin{center}
\caption{Components of TGO Trigger.\label{trigtab}}
\begin{tabular}{cc}
Trigger Name & Description \\
\hline
{${\it h}^{+}{\it h}^{-}$ } & ${\it M_{U}} \cdot {\it M_{D}}
 \cdot {\it S_{U}} \cdot {\it S_{D}}
 \cdot {\overline{\it NX1}} \cdot{\overline{\it NX3}}
 \cdot {\it H}$ \\
\hline
{${\mu}^{+}{\mu}^{-}$ } & ${\it M_{U}} \cdot {\it M_{D}}
 \cdot {\it S_{U}} \cdot {\it S_{D}}
 \cdot {\overline{\it NX1}} \cdot{\overline{\it NX3}}
 \cdot 2 {\it HX4} \cdot 2{\it HY4}$ \\
\hline
{${\it e}^{+}{\it e}^{-}$ } & ${\it M_{U}} \cdot {\it M_{D}}
 \cdot {\it S_{U}} \cdot {\it S_{D}}
 \cdot {\overline{\it NX1}} \cdot{\overline{\it NX3}}
 \cdot {\it E}$ \\
\hline
{${\it e}^{\mp}{\mu}^{\pm}$ } & ${\it M_{U}} \cdot {\it M_{D}}
 \cdot {\it S_{U}} \cdot {\it S_{D}}
 \cdot {\overline{\it NX1}} \cdot{\overline{\it NX3}}
 \cdot {\it e} \cdot {\it HX4} \cdot {\it HY4}$ \\
\hline
{${\it h}^{\mp}{\it e}^{\pm}$ } & ${\it M_{U}} \cdot {\it M_{D}}
 \cdot {\it S_{U}} \cdot {\it S_{D}}
 \cdot {\overline{\it NX1}} \cdot{\overline{\it NX3}}
 \cdot {\it h} \cdot {\it e}$ \\
\hline
{${\it h}^{\mp}{\mu}^{\pm}$ } & ${\it M_{U}} \cdot {\it M_{D}}
 \cdot {\it S_{U}} \cdot {\it S_{D}}
 \cdot {\overline{\it NX1}} \cdot{\overline{\it NX3}}
 \cdot {\it h} \cdot {\it HX4} \cdot {\it HY4}$ \\
\hline
{${\it h}^{\pm}{\it h}^{\pm}$ } & ${\it M_{LIKE}}
 \cdot ({\it S_{U}} + {\it S_{D}})
 \cdot {\overline{\it NX1}} \cdot{\overline{\it NX3}}
 \cdot {\it H}$ \\
\hline
{${\mu}^{\pm}{\mu}^{\pm}$ } & ${\it M_{LIKE}}
 \cdot ({\it S_{U}} + {\it S_{D}})
 \cdot {\overline{\it NX1}} \cdot{\overline{\it NX3}}
 \cdot 2 {\it HX4} \cdot 2{\it HY4}$ \\
\hline
{${\it e}^{\pm}{\it e}^{\pm}$ } & ${\it M_{LIKE}}
 \cdot ({\it S_{U}} + {\it S_{D}})
 \cdot {\overline{\it NX1}} \cdot{\overline{\it NX3}}
 \cdot {\it E}$ \\
\hline
{${\it e}^{\pm}{\mu}^{\pm}$ } & ${\it M_{LIKE}}
 \cdot ({\it S_{U}} + {\it S_{D}})
 \cdot {\overline{\it NX1}} \cdot{\overline{\it NX3}}
 \cdot {\it e} \cdot {\it HX4} \cdot {\it HY4}$ \\
\hline
{${\it h}^{\pm}{\it e}^{\pm}$ } & ${\it M_{LIKE}}
 \cdot ({\it S_{U}} + {\it S_{D}})
 \cdot {\overline{\it NX1}} \cdot{\overline{\it NX3}}
 \cdot {\it h} \cdot {\it e}$ \\
\hline
{${\it h}^{\pm}{\mu}^{\pm}$ } & ${\it M_{LIKE}}
 \cdot ({\it S_{U}} + {\it S_{D}})
 \cdot {\overline{\it NX1}} \cdot{\overline{\it NX3}}
 \cdot {\it h} \cdot {\it HX4} \cdot {\it HY4}$ \\
\end{tabular}
\end{center}
\end{table}

\begin{table}[hbt]
\begin{center}
\caption{Average number of protons on target and triggers per 23-sec spill.
\label{trigextab}}
\begin{tabular}{ccccc}
Run	  	& Protons on Target	& TFI	   	& TGO   	& TAP \\
\hline
900A-Be		& $3.6\times 10^{10}$& $2.8\times 10^{6}$& $1.7\times 10^{5}$& $8.7\times 10^{3}$  \\
900A-Au		& $3.3\times 10^{10}$& $8.3\times 10^{6}$& $1.0\times 10^{5}$	& $2.2\times 10^{4}$ \\
1000A-Au	& $2.6\times 10^{10}$ & $1.7\times 10^{6}$	& $1.0\times 10^{5}$	& $2.6\times 10^{4}$ \\
\end{tabular}
\end{center}
\end{table}

\begin{table}[hbt]
\begin{center}
\caption{Summary of data sets.} \label{datatab}
\begin{tabular}{cccc}
Data set	& Protons on target &
\raisebox{-.1ex}{ $AMON\cdot\overline{SB}$}	& TAPS \\
\hline
1000A-Au			& $7.2\times 10^{13}$ 
				& $3.9\times 10^{6}$	
				& $3.8\times 10^{8}$ \\
900A-Au				& $1.2\times 10^{13}$ 
				& $8.4\times 10^{5}$	
				& $8.7\times 10^{7}$ \\
900A-Au-Dedicated-Dilepton	& $6.6\times 10^{13}$ 
				& $3.4\times 10^{6}$	
				& $7.0\times 10^{7}$ \\
900A-Be				& $2.2\times 10^{14}$ 
				& $9.8\times 10^{5}$	
				& $7.3\times 10^{7}$ \\
\end{tabular}
\end{center}
\end{table}

\begin{table}[hbt]
\begin{center}
\caption{Lifetime Significance and impact parameter cuts 
for \dmumu.}
\label{mumucuts}
\begin{tabular}{lcccc}
	&900A-Au &900A-Au-Dedicated-Dilepton &900A-Be &1000A-Au\\
\hline
Lifetime Significance& 1.0 & 0.60 & 0.90 & 1.4 \\
Impact Parameter ($\mu$m)& 25.4 & 109 & 78.7 & 95.3 \\
\end{tabular}
\end{center}
\end{table}

\begin{table}[hbt]
\begin{center}
\caption{Lifetime Significance and impact parameter cuts for \dee.}
\label{eecuts}
\begin{tabular}{lcccc}
	&900A-Au &900A-Au-Dedicated-Dilepton &900A-Be &1000A-Au\\
\hline
Lifetime Significance & 0.80 & 0.90 & 0.60 & 0.80 \\
Impact Parameter ($\mu$m)& 48.3 & 82.6 & 29.2 & 55.9 \\
\end{tabular}
\end{center}
\end{table}

\begin{table}[hbt]
\begin{center}
\caption{Lifetime Significance and impact parameter cuts 
for \dmue.}
\label{muecuts}
\begin{tabular}{lcccc}
	&900A-Au &900A-Au-Dedicated-Dilepton &900A-Be &1000A-Au\\
\hline
Lifetime Significance& 1.5 & 1.3 & 1.0 & 1.4 \\
Impact Parameter ($\mu$m)& 190. & 97.8 & 90.2 & 55.9 \\
\end{tabular}
\end{center}
\end{table}

\begin{table}[hbt]
\begin{center}
\caption{Parameters of \dkpi\ normalization for \dmumu\ search.}
\label{mumunorm}
\begin{tabular}{lcccc}
Data set	&900A-Au	&900A-Be	&1000A-Au	&Dilepton\\
\hline
Mean Mass (GeV/$c^2$)& 1.865 & 1.863 & 1.867 & 1.866 \\
Mass Resolution, $\sigma$ (MeV/$c^2$) & 7.0 & 5.6	& 7.3 & 7.9 \\
Total \# of \dkpi\ decays&$606 \pm 56$ &$ 1161 \pm 110$ &$ 971 \pm 57$ & $2794 \pm 396$\\
\end{tabular}
\end{center}
\end{table}
 
\begin{table}[hbt]
\begin{center}
\caption{Parameters of \dkpi\ normalization for \dee\ search.}
\label{eenorm}
\begin{tabular}{lcccc}
Data set	&900A-Au	&900A-Be	&1000A-Au	&Dilepton\\
\hline
Mean Mass (GeV/$c^2$)&1.865 &1.863	& 1.866 & 1.865 \\
Mass Resolution, $\sigma$ (MeV/$c^2$)	& 6.1	& 5.9	& 6.5  & 6.2 \\
Total \# of \dkpi\ decays &$ 707 \pm 91$&$1563 \pm 154$&$1749 \pm 129$&$2867 \pm 466$ \\
\end{tabular}
\end{center}
\end{table}

\begin{table}[hbt]
\begin{center}
\caption{Parameters of \dkpi\ normalization for \dmue\ search.}
\label{muenorm}
\begin{tabular}{lcccc}
Data set	&900A-Au	&900A-Be	&1000A-Au	&Dilepton\\
\hline
Mean Mass (GeV/$c^2$)&1.865 & 1.864 & 1.867 & 1.865 \\
Mass Resolution, $\sigma$ (MeV/$c^2$)	& 6.0	& 6.5	& 7.1 	& 6.3 \\
Total \# of \dkpi\ decays&$255 \pm 21$ &$967 \pm 62$	&$1429 \pm 84$ &$2400 \pm 337$ \\
\end{tabular}
\end{center}
\end{table}

\begin{table}[hbt]
\begin{center}
\caption{Width of invariant-mass distribution for reconstructed dilepton events relative to that of the normalization signal, as determined by Monte Carlo.}
\label{relwidth}
\begin{tabular}{lccc}
	&\dmumu&\dee&\dmue\\
\hline
900A	&1.08&1.09&1.08\\
1000A	&1.09&1.08&1.07\\
\end{tabular}
\end{center}
\end{table}

\begin{table}[hbt]
\begin{center}
\caption{Efficiencies for \dkpi~decay.} \label{tabdkpieff}
\begin{tabular}{ccc}
	&900A	&	1000A \\
\hline
Geometric	& $(3.55 \pm 0.02) \times 10^{-3}$	& $(2.37
\pm 0.01) \times 10^{-3}$ \\
K decays	& $0.78 \pm 0.01$ 	& $0.82 \pm 0.01$ \\
Trigger	& $0.55 \pm 0.01$	& $0.58 \pm 0.01$ \\
Final cuts & $(7.12 \pm 0.04) \times 10^{-2}$	& $(9.35 \pm 0.05) 
\times 10^{-2}$ \\
\hline
Total	& $(1.08 \pm 0.02) \times 10^{-4} $	& $(1.05 \pm 0.02) \times
10^{-4}$ \\
\end{tabular}
\end{center}
\end{table}

\begin{table}[hbt]
\begin{center}
\caption{Efficiencies for \dmumu~decay.} \label{tabdmumueff}
\begin{tabular}{ccc}
	&900A	&	1000A \\
\hline
Geometric	& $(3.97 \pm 0.02) \times 10^{-3}$	& $(3.05 \pm 0.02) 
\times 10^{-3}$ \\
Trigger $\cdot$ ID $\cdot$ isolation	& $0.36 \pm 0.01$	& $0.50 \pm 0.01$	\\
Final cuts & $(7.30 \pm 0.04) \times 10^{-2}$	& $(9.85 \pm 0.05) 
\times 10^{-2}$ \\
\hline
Total	& $(1.05 \pm 0.02) \times 10^{-4} $	& $(1.51 \pm 0.02) \times
10^{-4}$ \\
\end{tabular}
\end{center}
\end{table}

\begin{table}[hbt]
\begin{center}
\caption{Efficiencies for \dee~decay.} \label{tabdeeeff}
\begin{tabular}{ccc}
	&900A	&	1000A \\
\hline
Geometric	& $(3.87 \pm 0.02) \times 10^{-3}$	& $(3.28 \pm 0.02)
 \times 10^{-3}$ \\
Trigger $\cdot$ ID	&$0.60 \pm 0.01$	&$0.60 \pm 0.01$ \\
Final cuts & $(7.15 \pm 0.04) \times 10^{-2}$	& $(10.0 \pm 0.05) \times
10^{-2}$ \\
\hline
Total	& $(1.65 \pm 0.03) \times 10^{-4}$	& $(1.98 \pm 0.04) \times
10^{-4}$ \\
\end{tabular}
\end{center}
\end{table}

\begin{table}[hbt]
\begin{center}
\caption{Efficiencies for \dmue~decay.} \label{tabdmueeff}
\begin{tabular}{ccc}
	&900A	&	1000A \\
\hline
Geometric	& $(4.18 \pm 0.02) \times 10^{-3}$	& $(3.42 \pm 0.02)
\times 10^{-3}$ \\
Trigger $\cdot$ ID&$0.36 \pm 0.01$ 	&$0.36 \pm 0.01$ \\
Final cuts & $(7.04 \pm 0.04) \times 10^{-2}$	& $(10.0 \pm 0.05) \times
10^{-2}$ \\
\hline
Total	& $(1.06 \pm 0.03) \times 10^{-4}$	& $(1.28 \pm 0.04) \times
10^{-4}$ \\
\end{tabular}
\end{center}
\end{table}

\begin{table}[hbt]
\begin{center}
\caption{Summary of results.} \label{results}
\begin{tabular}{ccccc}
Decay	&$\sum_{i} N_{i}\epsilon_{i}$ & Single-event 	& Limit from MC	& Cousins-Feldman \\
Mode	& 			& Sensitivity	& calculation & Method \\
\hline
\dmumu\ &    5830   & 7.34 $\times 10^{-6}$ & 1.56 $\times 10^{-5}$ & 1.65 $\times 10^{-5}$ \\
\dee\	&    11100  & 3.84 $\times 10^{-6}$ & 0.82 $\times 10^{-5}$ & 0.87 $\times 10^{-5}$ \\
\dmue\	&     5300  & 8.08 $\times 10^{-6}$ & 1.72 $\times 10^{-5}$ & 1.82 $\times 10^{-5}$ \\
\end{tabular}
\end{center}
\end{table}

\end{document}